\documentclass[useAMS,usenatbib]{mn2e}

\usepackage[english]{babel}
\usepackage[dvips]{graphicx}
\usepackage{lscape}

\title[The S0 Tully--Fisher relation]{The Tully--Fisher relation for S0 galaxies}
\author[A.G. Bedregal, A. Arag\'on-Salamanca and
  M.R. Merrifield]{A.G. Bedregal$^{1}$\thanks{E-mail:ppxapgg$@$nottingham.ac.uk},
  A. Arag\'on-Salamanca$^{1}$ and M.R. Merrifield$^{1}$  
\\
$^{1}$School of Physics and Astronomy, Centre of Astronomy and Particle
Theory, \\ University of Nottingham, University Park, Nottingham, NG7 2RD, UK}
\begin{document}

\date{Accepted ***. Received ***; in original form ***}

\pagerange{\pageref{firstpage}--\pageref{lastpage}} \pubyear{2002}

\maketitle

\label{firstpage}

\begin{abstract}
  We present a study of the local $B$- and $K_s$-band Tully--Fisher
  Relation (TFR) between absolute magnitude and maximum circular speed
  in S0 galaxies.  To make this study, we have combined kinematic
  data, including a new high-quality spectral data set from the Fornax
  Cluster, with homogeneous photometry from the RC3 and 2MASS
  catalogues, to construct the largest sample of S0 galaxies ever used
  in a study of the TFR.  Independent of environment, S0 galaxies are
  found to lie systematically below the TFR for nearby spirals in both
  optical and infrared bands.  This offset can be crudely interpreted
  as arising from the luminosity evolution of spiral galaxies that have faded
  since ceasing star formation. 

  However, we also find a large scatter in the TFR.  We show that most
  of this scatter is intrinsic, not due to the observational
  uncertainties.  The presence of such a large scatter means that the
  population of S0 galaxies cannot have formed exclusively by the
  above simple fading mechanism after all transforming at a single
  epoch.  
  To better understand the complexity of the transformation mechanism,
  we have searched for correlations between the offset from the TFR
  and other properties of the galaxies such as their structural
  properties, central velocity dispersions and ages (as
  estimated from line indices). For the Fornax Cluster data, the
  offset from the TFR correlates with the estimated age of the stars
  in the individual galaxies, in the sense and of the magnitude
  expected if S0 galaxies had passively faded since being converted
  from spirals.  This correlation implies that a significant part of the
  scatter in the TFR arises from the different times at which galaxies began
  their transformation.
\end{abstract}


\section{Introduction}

The Tully--Fisher relation (TFR; Tully $\&$ Fisher 1977) is one of the
most important physically-motivated correlations found in spiral galaxies. The correspondence
between luminosity and maximum rotational velocity ($V_{\rm max}$) is
usually interpreted as a product of the close relation between the
stellar and total masses of galaxies or, in other words, as the
presence of a relatively constant mass-to-light ratio in the local
spiral galaxy population (e.g. Gavazzi 1993, Zwaan et al. 1995). Such
a general property in spirals puts strong constraints on galaxy
formation scenarios (e.g. Mao et al. 1998, van den Bosch, 2000) and
cosmological models (e.g. Giovanelli et al. 1997, Sakai et al. 2000).
Also, the low scatter in the TFR (only $\sim 0.35$ mag in $I$-band,
according to Giovanelli et al. 1997, Sakai et al. 2000, Tully \&
Pierce 2000 and Verheijen 2001) permits us to use this tool as a good
distance estimator (e.g. Yasuda et al. 1997).

Attempts to ascertain whether S0 galaxies follow a similar TFR have
two main motivations.  First, if there is a TFR of S0s, it would prove
useful for estimating distances in the nearby universe, particularly
in clusters where S0s are very prevalent (Dressler 1980).  Second, and
more related to the present study, a possible scenario where S0
galaxies are the descendants of evolved spirals at higher redshifts
(Dressler 1980, Dressler et al. 1997) could leave traces of this
evolution in the observed TFR of S0s.  Different mechanisms have been
proposed as the channels for such evolution, like small mergers
(Schweizer 1986), gas stripping in later types (Gunn \& Gott 1972),
halo stripping (Bekki et al. 2002) and galaxy harassment (Moore et al.
1998). If this kind of picture is correct, it would be expected that
S0s retain some memory of their past as spirals, in particular through
their TFR, and perhaps even some clues as to which of the channels
they evolved down.

Only a few studies of the TFR for S0 galaxies can be found in the
literature. The first effort, made by Dressler \& Sandage (1983),
found no evidence for the existence of a TFR for S0 galaxies. However,
the limited spatial extent of their rotation curves, the inhomogeneous
photographic photometry employed and the large uncertainties in the
distances to their sample made it almost certain that any correlation
between luminosity and rotational velocity would be lost in the
observational uncertainties.

Fifteen years later, Neistein et al. (1999) explored the existence of
a TFR for S0s in the $I$-band, using a sample of 18 local S0s from the
field. Although some evidence for a TFR was uncovered in this study,
they also found a large scatter of $0.7$ magnitudes in the
relation, suggesting the presence of more heterogeneous evolutionary
histories for these galaxies when compared to spirals. Also, a
systematic shift $0.5$ magnitudes was found between their galaxies
and the relation for local spirals.

In two papers, Hinz et al. (2001, 2003), explored the $I$- and
$H$-band TFRs for 22 S0s in the Coma Cluster and 8 S0s in the Virgo
Cluster. By using cluster data, they avoided some of the errors that
arise from the uncertainty in absolute distances estimation. The
analysis of $I$-band data from the Coma Cluster revealed very similar
results to the study by Neistein et al. (1999), implying that the
larger scatter of the latter could not be attributed to distance
errors or the heterogeneous nature of the data.  In the $H$-band, an
even larger scatter of 1.3 magnitudes was found, but with a smaller
offset from the corresponding spiral galaxy TFR of only 0.2
magnitudes.  Interestingly, there was no evidence for any systematic
difference between the results for the Virgo and Coma Clusters,
despite their differences in richness and populations, implying that
these factors could not be responsible for the scatter in the
TFR.  Given the large scatter and small shift in the $H$-band TFR for
S0s compared to spirals, it was concluded that these
galaxies' properties are not consistent with what would be expected
for spiral galaxies whose star formation had been truncated; instead
they suggested that other mechanisms such as minor mergers are
responsible for the S0s' TFR.

By contrast, Mathieu, Merrifield \& Kuijken (2002) found in their
detailed dynamical modeling of six disk-dominated field S0s that these
galaxies obey a tight $I$-band TFR with a scatter of only 0.3
magnitudes, but offset from the spiral galaxy TFR by a massive 1.8
magnitudes.  The authors therefore concluded that these objects were
consistent with being generated by passively fading spirals that had
simply stopped producing stars.  This result does not appear to be
consistent with the previous studies, although it should be borne in
mind that the galaxies in this study were selected to be
disk-dominated, so they morphologically resemble spiral galaxies more
than those in other work.  In addition, their field locations means
that they are less likely to have had their evolution complicated by
mergers.  It is therefore possible that these S0s really are just
passively-fading spirals where those in clusters have led more
complicated lives.  

As can be seen, there is no general consensus as to either the scatter
or the shift in the TFR for S0 galaxies when compared to spirals, and
so no agreement as to their interpretation.  We therefore revisit the
TFR of S0 galaxies in this paper, adding our new spectral dataset from
the Fornax Cluster to the existing results from literature.  In
addition to the new homogeneous spectral data set, we can also take
advantage of the 2MASS photometry for these galaxies (Jarrett et
al. 2003), to obtain consistent infrared magnitudes and structural
parameters for all the galaxies.  With this more systematic study of
S0s, it is to be hoped that we can avoid any past problems that may
have arisen due to the heterogeneous nature of the available data, to
reveal the underlying physics that dictates the TFR in S0 galaxies.

The remainder of the paper is laid out as follows.  In
Section~\ref{sec:data}, we describe the different samples of S0s used
in this study, and the data that we have adopted.
Section~\ref{sec:results} presents the main results and then discusses
their implications. Finally in Section~\ref{sec:conc} our conclusions
are summarised.

\section{The data}\label{sec:data}
\subsection {Kinematics}
To build the TFR of local S0 galaxies, we have collated the data on
their kinematics from four previous studies. These works are: Neistein
et al. (1999), hereafter N99; Hinz et al. (2001, 2003), hereafter H01
and H03, respectively; and Mathieu, Merrifield \& Kuijken (2002), hereafter
M02. From the sample of N99, we exclude the galaxy NGC\,4649 as it has a low
degree of rotational support and it presents evidence of interaction
with a neighbouring system. From H01, the Sab spiral galaxy IC\,4088
was also excluded.  To these data, we have added the observations that
we have recently obtained using the VLT of S0 galaxies in the Fornax
Cluster (Bedregal, Arag\'on-Salamanca \& Merrifield 2006, hereafter Paper I);
of the galaxies observed in this cluster, seven are rotationally supported
systems, so are suitable for this study.  This combined data set
provides us with 60 S0 galaxies with measured kinematics, the largest
sample yet used in a study of the S0 TFR.  The collated kinematic data
values for the maximum rotation speed, $V_{\rm max}$ (for all the sample), and
the central velocity dispersion, $\sigma_0$ (for 51 galaxies of the sample), are listed in Table~A1.

\subsection{Photometry}
In the present study, we have adopted $K_s$-band photometry from the Two
Micron All Sky Survey (2MASS, Jarrett et al. 2003) and $B$-band
photometry from the Third Reference Catalogue of Bright Galaxies (RC3,
de Vaucouleurs et al. 1991). Of the complete sample of 60 galaxies,
photometry in $K_s$-band is available for all objects, while 54
galaxies have photometry in $B$-band.

In order to convert these data to absolute magnitudes, we need
distances to all objects in the sample.  In the clusters, we adopted
distance moduli of 31.35 for members of the Fornax Cluster (Madore et
al. 1999), a redshift of 0.0036 for the Virgo Cluster members (Ebeling et al.
1998), and a redshift of 0.0227 for Coma Cluster galaxies (Smith et al. 2004).  For the field
galaxies from N99, the distance moduli of Tonry et al. (2001) were
used. Finally, for the M02 sample, redshifts from the NASA/IPAC
Extragalactic Database were used where available. For two of their
galaxies (NGC\,1611 and NGC\,2612), an estimate of the distance was
calculated by the authors, using the systemic velocity derived from
their spectra. A Hubble constant of $70\,$km$\,$s$^{-1}\,$Mpc$^{-1}$
was adopted in converting redshifts into distances.  Galactic
extinction corrections were calculated using the Schlegel, Finkeiner
\& Davis (1998) reddening curve description, $A_b^{\rm \lambda}=R_{\rm
\lambda}E(B-V)$, where $R_{\rm \lambda}$=4.32 and 0.37 for $\lambda=B$
and $K_s$, respectively. In the $B$-band, we applied the
$k$-correction from Poggianti et al. (1997); no correction was applied
for the $K_s$-band, as it is negligible.  No internal extinction
correction was applied to the apparent magnitudes; there is no
definitive study on the internal extinction of S0 galaxies, but the
apparent lack of dust in these systems suggests that such a correction
would be very small.  The resulting absolute magnitudes in $K_s$ and $B$-bands
are listed in Tables~A1 and A2, respectively.

In addition to the absolute magnitudes, we derived structural
parameters from the spatially-extended photometry available for these
galaxies.  The data used were the publicly-available ``postage stamp''
images in the $K_s$-band from 2MASS. Bulge-plus-disk models were fitted
directly to these images using GIM2D (Simard et al. 2002), with a
S\'ersic law adopted for the bulge distribution, and an exponential
for the disk.  In this way, we derived values for the bulge effective
radius, $R_e$, its S\'ersic index, $n$, the disk scalelength, $R_d$,
the half-light radius, $R_{\rm half}$, the bulge-to-total fraction,
$B/T$, and the galaxy inclination $i$.  In a few cases, the derived
bulge scale length was found to be smaller than the seeing ($\sim 2.5$
arcsec). In those cases, the structural parameters are not well
constrained by the observations, so the values were excluded.  The
structural parameters derived for the remaining 48 galaxies are listed
in Table~A1.

\subsection{Line indices, ages and metallicities}
We have full spectral data for 7 Fornax Cluster galaxies from which we can
extract Lick indices. A detailed description of the line indices calculation
is given in our forthcoming study of the stellar populations of these galaxies
(Bedregal et al. 2007, hereafter Paper III). Briefly, we convolved the spectral
bins with appropriate gaussians (including galactic and instrumental
dispersions) in order to achieve the $3\, \rm \AA$ resolution of Bruzual \&
Charlot (2003) (hereafter BC03) simple stellar populations models. The indices
$H\beta$, $Mgb$, $Fe\lambda5270$ and $Fe\lambda5335$ were measured within
$R_e$/8 (``Central'' values), between 0.75 and 1.25 $R_e$ (``$1\,R_e$''
values) and between 1.5 and 2.5 $R_e$ (``$2\,R_e$'' values).The resulting
values are presented in Tables~A2 and A3. 

 Central line indices for a handful of objects in our sample can be found in
 the literature (Fisher et al. 1996, Terlevich et al. 2002, Denicolo et
 al. 2004)  mainly corresponding to field S0s from  the N99
 subsample. Unfortunately, these datasets mainly include the brightest objects
 of the sample, so it is difficult to make meaningful comparisons with the
 fainter galaxies from Fornax. Also, differences between the two samples could
 arise because of the superior quality of Fornax data, effect which is
 difficult to quantify. In consequence, we will focus our analysis of the line
 indices, ages and metallicities on the data from the Fornax Cluster only.

From Fornax galaxies' indices, one can derive measures of the
luminosity-weighted ages and metallicities. To this end, we have used the
simple single-age stellar population models of BC03 to translate the measured
line indices into estimates of age and metallicity.  As a check on the
uncertainties inherent in this process, we have repeated the
calculations using the $Mgb$ index and the combined indices
$\langle Fe \rangle$ (Gorgas, Efstathiou \& Arag\'on-Salamanca 1990)
and $[MgFe]'$ (Gonz\'alez 1993; Thomas, Maraston \& Bender 2003) as the
metallicity-sensitive index. Solar abundance ratios were assumed for the
models. The resulting ages, metallicities and their uncertainties (which
include the effects of covariance between the two parameters) are also shown
in Tables~A2 and A3. 

\section[]{Results and Discussion}\label{sec:results}
\begin{figure}
\begin{center}
\includegraphics[scale=0.4]{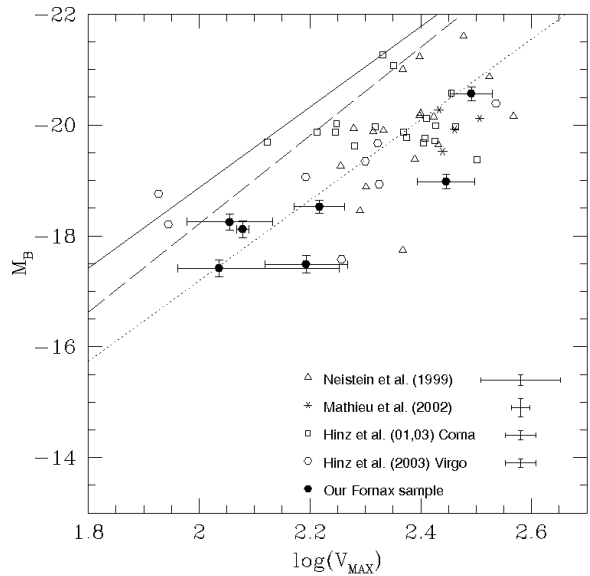}
\end{center}
\caption{\small $B$-band Tully--Fisher relation of S0 galaxies. Solid
  line represents the TFR of spiral galaxies from Tully \& Pierce
  (2000); dashed line represents the spiral TFR by
  Sakai et al. (2000); dotted line is the best fit to the S0
  data-points using the slope from Tully \& Pierce (2000). The error
  bars in the right down corner correspond to the median value for
  each subsample, while for the Fornax Cluster data we plot the errors
  for each data-point.}\label{fig:TFRB}
\end{figure}

\begin{figure}
\begin{center}
\includegraphics[scale=0.4]{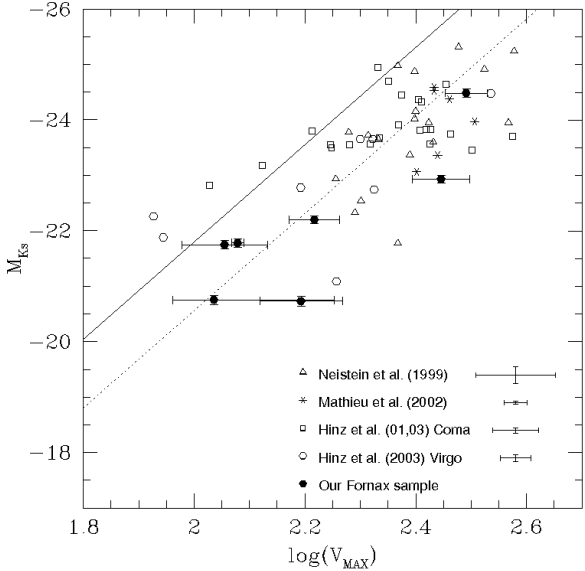}
\end{center}
\caption{\small $K_s$-band Tully--Fisher relation of S0
  galaxies. Solid line represents the TFR of spiral galaxies from
  Tully \& Pierce (2000); dotted line is the best fit to the S0
  data-points using the slope from Tully \& Pierce (2000).The error
  bars in the right down corner correspond to the median value for
  each subsample, while for the Fornax Cluster data we plot the errors
  for each data-point.}\label{fig:TFRKs}
\end{figure}

The basic result of this analysis is presented in the Tully--Fisher
plots of rotation velocity versus absolute magnitude shown in
Figures~\ref{fig:TFRB} and \ref{fig:TFRKs} (for the $B$- and
$K_s$-band respectively).  In these plots, the solid line represents
the TFR of spirals in local clusters found by Tully \& Pierce (2000,
hereafter TP00), shifted by $-0.207$ magnitudes in order to be
consistent with the adopted Hubble constant of
$H_0=70\,$km$\,$s$^{-1}\,$Mpc$^{-1}$. The long-dashed line in the
$B$-band represents the TFR of cluster spirals by Sakai et al. (2000,
hereafter Sak00). The difference between these lines give an
indication of the remaining systematic uncertainty in the spiral
galaxy TFR with which we seek to compare the S0s.


\subsection{Shift between the spiral and S0 TFRs}\label{sec:offset}
The first point that is immediately clear from Figures~\ref{fig:TFRB} and
\ref{fig:TFRKs} is that, as found by previous authors, whichever spiral
galaxy TFR we adopt, the S0s lie systematically below it.  It is also
interesting to note that this result holds true for the S0 data from
all environments, from the poorest field objects to fairly rich
clusters, so it is clearly a very general phenomenon.  We therefore
now seek to quantify and model the possible origins of such an offset.

\subsubsection{Observational results}
One problem in trying to quantify the offset in the TFR is that the
incompleteness in magnitude of the data presented in
Figures~\ref{fig:TFRB} and \ref{fig:TFRKs} will bias a conventional fit.  We
therefore adopt the approach of Willick (1994), which involves fitting
the inverse function,
\begin{equation}
log(V_{\rm max})=a+b M_{\rm \lambda}, \mbox{$\hspace{10 mm}$ where $\lambda$ =
  $B$ or $K_s$},
\end{equation}
to minimise this source of bias.  The slope $b$ is fixed to match the
slope for the spiral galaxy TFR, and $a$ is varied to find the
least-squares fit between this function and the data, with each point
$i$ weighted by
\begin{equation}\label{eq:wi}
w_i=\frac{1}{\sigma_i^2},
\end{equation}
where
\begin{equation}
\sigma_i^2= \sigma_{\rm log(Vmax),i}^2 + b^2\sigma_{\rm M_{\lambda},i}^2 +
\sigma_{\rm int}^2,
\end{equation}
to account for the uncertainty in the measured maximum velocity,
$\sigma_{\rm log(Vmax),i}$, and that in the absolute magnitude,
$\sigma_{\rm M_{\lambda},i}$.  The quantity $\sigma_{\rm int}$ is
set to quantify the intrinsic scatter in the relation such that the
reduced $\chi^2$ of the fit comes out at unity; this procedure is
discussed in more detail in Section~\ref{Sec:scatter}.

Setting $b$ equal to the inverse of the spiral TFR slope determined by
TP00, we can determine the zero-point parameter $a$ and hence the
offset in magnitudes from the TP00 TFR relations in the $B$- and
$K_s$-band. The resulting best-fit lines are shown dotted in
Figures~\ref{fig:TFRB} and \ref{fig:TFRKs}.  The offsets from the TP00
relations are 
\begin{equation}\label{eq:TPoffB}
\Delta M_{B,TP00} = -1.7 \pm 0.4
\end{equation}
and
\begin{equation}\label{eq:TPoffKs}
\Delta M_{K_s,TP00} = -1.2 \pm 0.4,
\end{equation}
where the quoted error includes the uncertainty in zero point of both
the S0 and spiral TFRs.  To test the robustness of this result against
the uncertainty in the spiral TFR, we repeated the analysis using the
Sak00 $B$-band relation to fix the slope and measured the offset from
their relation.  This analysis resulted in an offset of
\begin{equation}\label{eq:SakoffB}
\Delta M_{B,\rm Sak00} = -1.3 \pm 0.1,
\end{equation}
within the errors of the previous analysis but somewhat smaller.
Sak00 did not publish a $K_s$-band TFR, but the parallel nature of the
Sak00 and TP00 TFRs in Figure~\ref{fig:TFRB} suggest that most of the
difference arose from a zero-point shift due to a different
distance-scale calibration.  We might therefore extrapolate from the
Sak00 $B$-band results on the assumption that $\Delta M_{B,TP00} -
\Delta M_{K_s,TP00} \approx \Delta M_{B,\rm Sak00} - \Delta M_{K_s,\rm
Sak00}$, to infer a corresponding $K_s$-band offset of
\begin{equation}\label{eq:SakoffKs}
\Delta M_{K_s,\rm Sak00} = -0.8 \pm 0.4.
\end{equation}

These new estimates for the offset between spiral and S0 TFRs tend to
lie toward the upper end of earlier estimates, mainly because of the
more recent refinements in the calibration to the spiral galaxy TFR
that we have included in this analysis.  However, even neglecting
these systematic changes, the values obtained here lie within the
range of offsets in the TFR suggested by previous studies, implying
that this quantity can be fairly reliably determined, particularly
with the larger and more homogeneous data set presented here.

\subsubsection{Interpretation}\label{sec:shiftint} 
The most natural interpretation of the offset between the S0 and
spiral TFRs is that it represents a simple fading as a spiral galaxy's
star formation is truncated when it mutates into an S0.  One
complication in attempting to quantify this scenario is that one needs
to know what luminosity the spiral galaxy started at in this
evolution.  However, observations out to redshifts beyond unity seem to show
that there has been essentially no evolution in either the slope or
zero-point of the spiral galaxy TFR in optical or infrared wavebands
over this period (Vogt et al. 1997, Conselice et al. 2005, Bamford et
al. 2006; although see Rix et al. 1997 and B\"ohm et al. 2004 for alternative
views). This lack of evolution means that the starting point for
fading spirals is the same spiral galaxy TFR that we see today, and the
offset between the nearby galaxy spiral and S0 TFRs does provide an
accurate measure of the degree to which the S0 galaxies must have
faded if this picture is correct.  

To see if this scenario is plausible and to quantify the timescales
involved, we have used the BC03 synthesis models
to calculate the fading of a stellar population that started with a
constant star formation rate of $1\,M_{\odot}$yr$^{-1}$ for 5 Gyrs,
then stopped.  The stellar population was assumed to have solar
metallicity and the initial mass function of Chabrier (2003).
By matching the decrease in luminosity to the observed shifts between
spiral and S0 TFRs, we can determine how long ago the truncation in
star formation must have occurred to be consistent with our
observations.  Using the TP00 offsets in the $B$- and $K_s$-bands
given in equations~\ref{eq:TPoffB} and \ref{eq:TPoffKs}, we
thus obtain times since truncation of
\begin{equation}
\tau_{\rm B,TP00}^{\rm trunc} = 1.6^{+1.2}_{-0.7} \mbox{Gyrs}
\end{equation}
and
\begin{equation}
\tau_{\rm K_s,TP00}^{\rm trunc} = 6.2^{+\infty}_{-3.1} \mbox{Gyrs,} 
\end{equation}
respectively.  These values are somewhat inconsistent with each other,
but this may just reflect the calibration of the spiral TFR in TP00;
if we instead use the values that we infer from the Sak00 calibration
(equations~\ref{eq:SakoffB} and \ref{eq:SakoffKs}), we obtain more
consistent timescales since truncation of
\begin{equation}
\tau_{\rm B,Sak00}^{\rm trunc} = 0.66^{+0.63}_{-0.30} \mbox{Gyrs}
\end{equation}
and
\begin{equation}
\tau_{\rm K_s,Sak00}^{\rm trunc} = 1.1^{+1.6}_{-1.0} \mbox{Gyrs}. 
\end{equation}
The shorter timescale simply reflects the smaller offset in the TFR
that the Sak00 calibration implies.  However, this timescale is if
anything worryingly short: it would be quite a coincidence if we
are living at an epoch so close to the point at which all these
galaxies transformed from spirals into S0s.  

One possible resolution is that the star formation history in
transforming a spiral into an S0 might be somewhat more complex.
Indeed, Poggianti et al. (1999) have suggested that cluster galaxies
with k+a/a+k spectra observed at intermediate redshifts are the best
candidates for S0 progenitors because of their spectro-photometric
characteristics and the predominance of disk-dominated morphologies.
Since such spectra are usually identified with post-starburst
galaxies, it would seem that spiral galaxies may undergo a ``last
gasp'' burst of star formation when they start their transitions into
S0s.  To investigate this possibility, we added a burst of star
formation, amounting to 10\% of the total stellar mass, to the above
truncated star formation model.  Repeating the comparison between the
luminosity of this model as predicted by the BC03 population synthesis code
and the observed offsets in the TFR resulted in estimates for the time since
the burst and truncation of
\begin{equation}
\tau_{\rm B,TP00}^{\rm burst} = 2.3^{+1.3}_{-0.8} \mbox{Gyrs},
\end{equation}
\begin{equation}
\tau_{\rm K_s,TP00}^{\rm burst} = 7.8^{+\infty}_{-3.7} \mbox{Gyrs} 
\end{equation}
using the TP00 calibration, and
\begin{equation}
\tau_{\rm B,Sak00}^{\rm burst} = 1.1^{+0.7}_{-0.4} \mbox{Gyrs,}
\end{equation}
\begin{equation}
\tau_{\rm K_s,Sak00}^{\rm burst} = 1.7^{+1.9}_{-1.1} \mbox{Gyrs} 
\end{equation}
using the Sak00 values.  As might be expected, the inclusion of a
starburst increases the age of the stellar population compared to
those found in the truncation model. These values are therefore a
little more comfortable in terms of the timescales involved, and, at
least using the Sak00 calibration, produce fairly consistent results
between different wavebands.  

However, there is a limit to how far it is worth pursuing these simple
models, as it is extremely unlikely that all S0 galaxies will have
undergone the same star formation history.  Indeed, the large scatter
apparent in the points in Figures~\ref{fig:TFRB} and \ref{fig:TFRKs}
indicates a relatively heterogeneous history for these systems: the
average evolution may be as described above, but each galaxy has its
own story to tell.  We therefore now look in more detail at the
scatter in the S0 TFR, in an attempt to quantify it and explore its
origins.

\subsection{The Scatter in TFR of S0 galaxies}\label{Sec:scatter}

\subsubsection{Observational results}
As outlined in the previous section, we estimate the intrinsic scatter
in the TFR, $\sigma_{\rm int}$, during the fitting process by varying
its value in the weights of equation~\ref{eq:wi} until the reduced
$\chi^2$ of the fit, 
\begin{equation}\label{eq:chi2r}
\chi_r^2 = \frac{1}{n-2} \sum_i\Big(\frac{\log(V_{\rm max,i})- a - bM_{\rm
    \lambda,i}}{\sigma_i}\Big)^2, 
\end{equation}
was equal to unity.  The inverse slope of the TFR, $b$, was set to the
value appropriate to either the TP00 or the Sak00 spiral TFR (the
values are so similar that it made no substantial difference to the measured
scatter), while the zero-point $a$ was allowed to vary.  The presence
of variables in both numerator and denominator of equation
\ref{eq:chi2r} mean that the fit is no longer linear, but we found
that it could be robustly performed by a simple iterative procedure in
which at the $j$th iteration the estimate for $\sigma_{\rm int}$ was
updated such that
\begin{equation}
\sigma_{\rm int,j}^2 = \sigma_{\rm int,j-1}^2\times\chi_r^{2\alpha}.
\end{equation}
By setting the convergence parameter $\alpha$ to $2/3$, we found that
the iterative solution was well behaved and converged in no more than
15 iterations.  For comparison, we also calculated a weighted total
scatter using the formula
\begin{equation}\sigma_{\rm tot}^2 = 
        \frac{\sum_iw_i(\log(V_{max,i})-a-bM_{\rm \lambda,i})^2}{\sum_iw_i}. 
\end{equation}

For the $B$-band TFR of S0s (using TP00 slope), we thus derived
\begin{equation}
\begin{array}{rcl}
\sigma_{\rm tot,B} &=& 0.88 \pm 0.06 \mbox{ mag}, \\
\sigma_{\rm int,B} &=& 0.78 \pm 0.06 \mbox{ mag},
\end{array}
\end{equation}
while for $K_s$-band data (using TP00 slope) we found
\begin{equation}
\begin{array}{rcl}
\sigma_{\rm tot,K_s} &=& 0.98 \pm 0.06 \mbox{ mag}, \\
\sigma_{\rm int,K_s} &=& 0.87 \pm 0.06 \mbox{ mag}.
\end{array}
\end{equation}
These values imply that $\approx$ 90$\%$ of the observed scatter in
Figures~\ref{fig:TFRB} and \ref{fig:TFRKs} cannot be explained by the
known observational uncertainties in $M_{\rm \lambda}$ and log($V_{\rm
max}$).  These results seem to be bracketed by previous estimates: H03
report a scatter in the $H$-band TFR of 1.18 magnitudes in the Coma
Cluster and 1.33 in the Virgo Cluster, whereas previous $I$-band
estimates of scatter have been lower at 0.68 magnitudes (N99) and 0.82 magnitudes
(H03).  However, these previous studies are not directly comparable to
the current estimates, because of their different wavebands and
because they did not use the more robust inverse-fitting approach
adopted here.  In addition, it is not entirely clear whether the
previous estimates have been corrected for measurement error,
particularly in the uncertain measurement of $\log(V_{\rm max})$.

\subsubsection{Interpretation}\label{sec:scatterint}
Perhaps the simplest possible explanation for the large value
$\sigma_{\rm int}$ is that the errors in the observed quantities
plotted in Figures~\ref{fig:TFRB} and \ref{fig:TFRKs} have been
underestimated, and have been erroneously attributed to the intrinsic
scatter.  We therefore begin by considering possible additional
sources of uncertainty in the data.

The adopted fluxes for the sample galaxies, particularly the infrared
2MASS data, form a uniformly-measured reliable set, so it is unlikely
that much of the scatter in the TFR could arise from errors in these
data.  However, their transformation into absolute magnitudes could
introduce some uncertainty.  For example, we have not applied any
correction to the magnitudes due to the internal extinction of these
systems.  Although such corrections are believed to be small in the
relatively dust-free environment of an S0, they might still be
non-negligible.  However, if such an effect were distorting the
results, we would expect it to have a much stronger effect in the
$B$-band than in the $K_s$-band; the very similar values for
$\sigma_{\rm int}$ derived in both bands imply that extinction is not
a significant issue.  Similarly, it is possible that unaccounted
errors in the adopted distances to the sample galaxies could
contribute to the scatter in absolute magnitudes.  However, although
such an error could offset all the points derived from S0s in a single
cluster, it would not increase their scatter.  The lack of offsets
between the Virgo, Coma and Fornax Cluster data and their similar
large scatters in Figures~\ref{fig:TFRB} and \ref{fig:TFRKs} imply
that such errors are not significant.

The measurement of $\log(V_{\rm max})$ is more challenging.  The
measured Doppler shift in the stellar component is only indirectly
related to this quantity.  First, since we only measure the
line-of-sight component of velocity, a correction must be applied for
inclination.  However, it is in the nature of identifying S0 galaxies
that only those relatively close to edge-on are classified as such
(Jorgensen \& Franx 1994): as Table~A1 confirms, the vast majority of
galaxies in the sample are at inclinations close to 90 degrees, so the
corrections are relatively small.  Second, the stars in S0s do not
follow perfectly circular orbits, so the measured rotation velocity
will differ from the circular speed by a quantity termed the
``asymmetric drift'' (Binney \& Tremaine 1987).  For the Virgo, Coma
and Fornax Cluster data, we have adopted the approach of N99 that uses
the measured random velocities to correct for the asymmetric drift;
although this relatively simple correction may contain systematic
uncertainties, it is unlikely to increase the scatter significantly.
It is also notable that the new Fornax Cluster kinematic data used
here reach to larger radii than the earlier samples.  Since the size
of the asymmetric drift correction decreases with radius, we would
expect any scatter induced by it to be smaller for these larger-radii
data, but no differences are discernible.  Finally, we note that M02
used a more sophisticated dynamical modeling technique that removed
the need for any asymmetric drift correction.  Although they claimed a
resulting decrease in the scatter in the TFR, the better photometry
presented in this paper suggests that this is only marginally the
case.  Irrespective of the way that $\log(V_{\rm max})$ is derived,
there seems to be a sizeable residual scatter in the TFR.

We therefore now consider possible astrophysical origins for the
scatter in the TFR.  In this context, it is notable that the scatter
that we derive, although much larger than that seen in the TFR of
nearby spirals ($\sim 0.4$ magnitudes; TP00, Sak00, Verheijen 2001),
is similar to the scatter observed in the TFR of galaxies at
higher redshift.  Bamford et al. (2006) obtained similar
values for $\sigma_{\rm int}$ for their sample of high-redshift
spirals in the $B$-band, while Conselice et al. (2005) found a similar
scatter in their $K$-band TFR of spirals at redshifts between 0.2 and
1.2.  This larger scatter in the spiral TFR at higher redshift has
been attributed to variations in the star formation rate in these
systems due to more frequent interactions with other galaxies, gas
clouds and cluster environment, as well as the less relaxed dynamical
state of these systems (e.g. Shi et al. 2006, Flores el at. 2006).  It
is therefore possible that the scatter in the S0 TFR was in some sense
imprinted into these systems while they were still ``normal'' spiral
galaxies at higher redshift, and that scatter has remained frozen into
their TFR as they have evolved more gently to the present day.
Alternatively, the scatter could be a signature of the transformation
process itself.  For example, if the transition from spiral to S0
began over a range of times, then different galaxies will have faded
by different amounts, leading to the observed scatter in the relation.
As we have seen in Section~\ref{sec:offset}, the mean TFR will shift
downward by more than a magnitude on billion-year timescales, so a
spread in start times could explain the ultimate scatter.  As a
further complication, subsequent minor mergers might induce extra late
bursts of star formation, or they could kill off the galaxy's star
formation a little sooner, further scattering the relation.  In order
to try to distinguish between these possibilities, we now investigate
whether the offset of individual S0s from the spiral galaxy TFR
correlates with any of their other astrophysical properties.

\begin{figure}
\begin{center}
\includegraphics[scale=0.4]{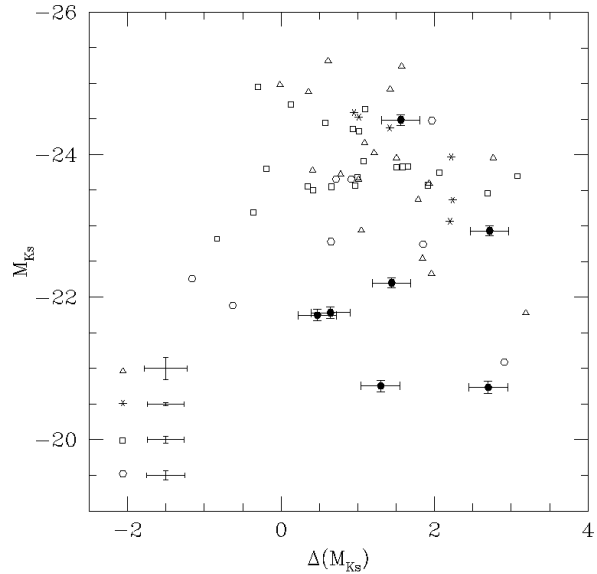}
\end{center}
\caption{Plot of infrared absolute magnitude, $M_{\rm K_s}$, against
  offset from the TFR, vs $\Delta M_{\rm K_s}$. In the lower left corner,
  the median uncertainties for the N99, M02, H01 and H03 subsamples
  are shown, while errors for the individual Fornax Cluster data are
  shown.}\label{fig:MK_v_DMK}
\end{figure}

\subsection{Correlations with other parameters}
\subsubsection{Structural parameters}
In searching for correlations between the offset from the infrared
TFR, $\Delta M_{K_s}$, and other structural parameters, one concern is
that the offset may not be the driving variable.\footnote{Here we
  discuss the results for the $K_s$-band data, since the structural parameters
  were calculated from the same photometry and the infrared dataset is
  somewhat more uniform and reliable.  However, similar results are
  found if the analysis is performed using the optical $B$-band data.}
In particular, residual bias in the fitting of the TFR to
Figures~\ref{fig:TFRB} and \ref{fig:TFRKs} could induce a correlation
between $\Delta M_{K_s}$ and the absolute magnitude of the galaxies,
$M_{K_s}$.  Since it is well know that other properties such as
galaxies' sizes correlate with their magnitudes, then such a bias in
the TFR fit would also induce correlations with $\Delta M_{K_s}$.  In
practice, Figure~\ref{fig:MK_v_DMK} shows that there is only the
slightest hint of an anti-correlation between the derived values
$\Delta M_{K_s}$ and $M_{K_s}$, so this potential source of bias has
clearly been dealt with reasonably effectively.

\begin{figure}
\begin{center}
\includegraphics[scale=0.4]{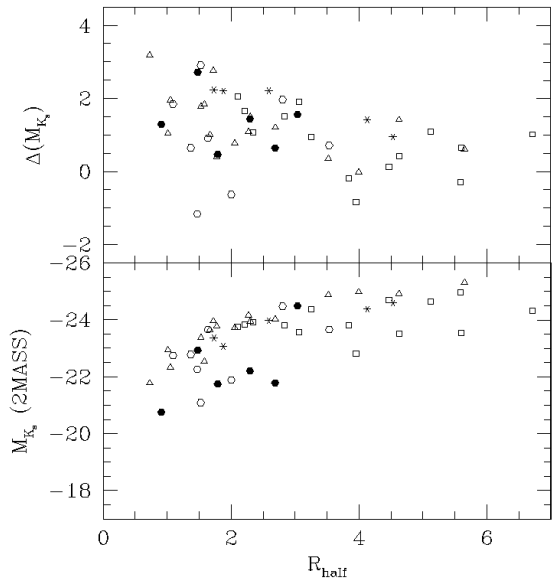}
\end{center}
\caption{$M_{\rm K_s}$ and $\Delta M_{\rm K_s}$ versus half-light
radius.}\label{fig:v_rhalf}
\end{figure}

\begin{figure}
\begin{center}
\includegraphics[scale=0.4]{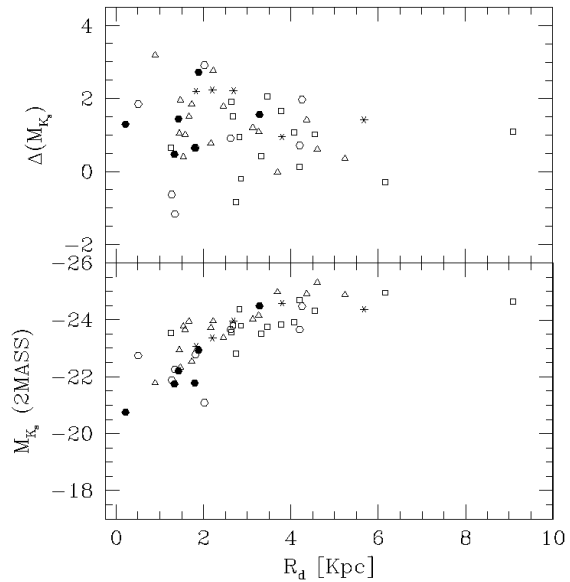}
\end{center}
\caption{$M_{\rm K_s}$ and $\Delta M_{\rm K_s}$ versus disk
scalelength.}\label{fig:v_rdisk}
\end{figure}

\begin{figure}
\begin{center}
\includegraphics[scale=0.4]{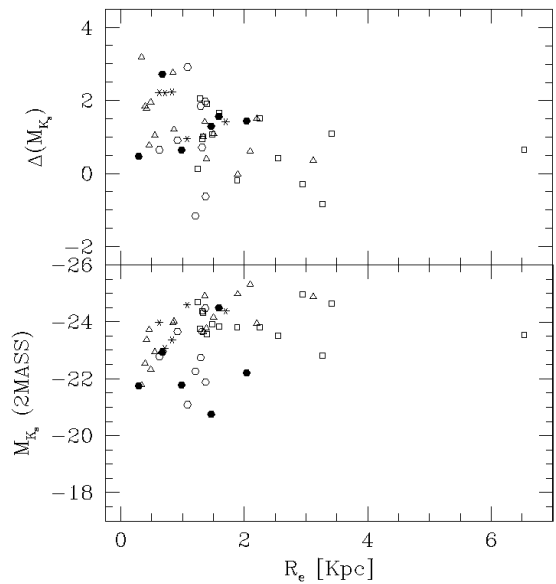}
\end{center}
\caption{$M_{\rm K_s}$ and $\Delta M_{\rm K_s}$ versus bulge effective
radius.}\label{fig:v_reff}
\end{figure}

\begin{figure}
\begin{center}
\includegraphics[scale=0.4]{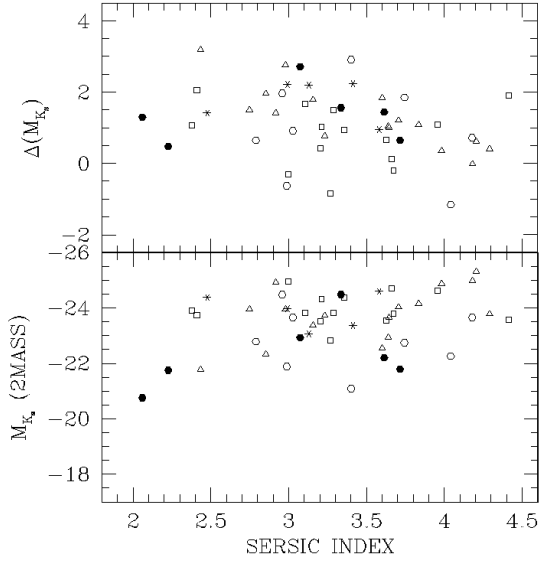}
\end{center}
\caption{$M_{\rm K_s}$ and $\Delta M_{\rm K_s}$ versus S\'ersic Index of the
bulge.}\label{fig:v_sersic}
\end{figure}

\begin{figure}
\begin{center}
\includegraphics[scale=0.4]{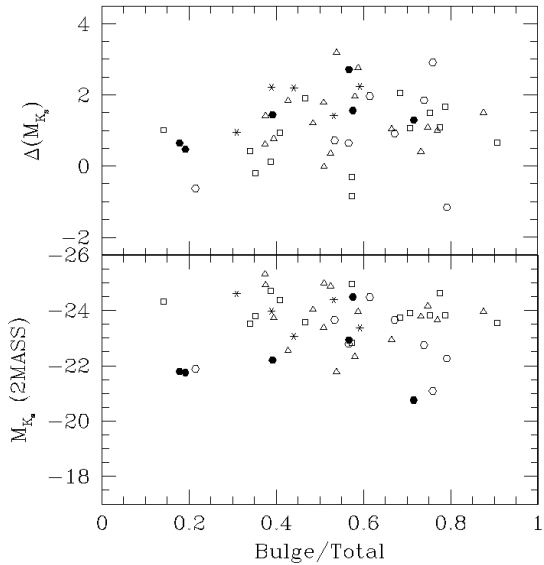}
\end{center}
\caption{$M_{\rm K_s}$ and $\Delta M_{\rm K_s}$ versus bulge-to-total
  luminosity ratio.}\label{fig:v_BoverT}
\end{figure}
Figures 4 -- 9 show the correlations between the various structural properties
of the S0 galaxies (as presented in Table~A1 in the appendices), and their
absolute magnitudes and offset from the spiral galaxy TFR in $K_s$-band.  The
symbols used in these figures are the same as in Figures~\ref{fig:TFRB} and
\ref{fig:TFRKs}.  Some of the resulting correlations are fairly trivial.  For
example, the absolute magnitude correlates quite well with the size of the
galaxy, as characterised by its half-light radius (see
Figure~\ref{fig:v_rhalf}). It is, however, interesting to note that this
correlation is mainly driven by the size of the disks in these systems (see
Figure~\ref{fig:v_rdisk}), and is almost uncorrelated with the
photometric properties of the bulge (see Figures~\ref{fig:v_reff} and
\ref{fig:v_sersic}).  Similarly, as Figure~\ref{fig:v_BoverT} shows,
the magnitudes do not seem to be systematically affected by how much
of the total luminosity is in the bulge or disk components. However, we also
notice that cluster galaxies are mainly responsible of the
scatter in Figures~\ref{fig:v_rhalf} and \ref{fig:v_reff}, while field S0s
tend to follow tighter trends in both diagram. However, such a difference
between cluster and field galaxies could be related to different selection
criteria, so any further interpretation would be excessive. 
  
\begin{figure}
\begin{center}
\includegraphics[scale=0.4]{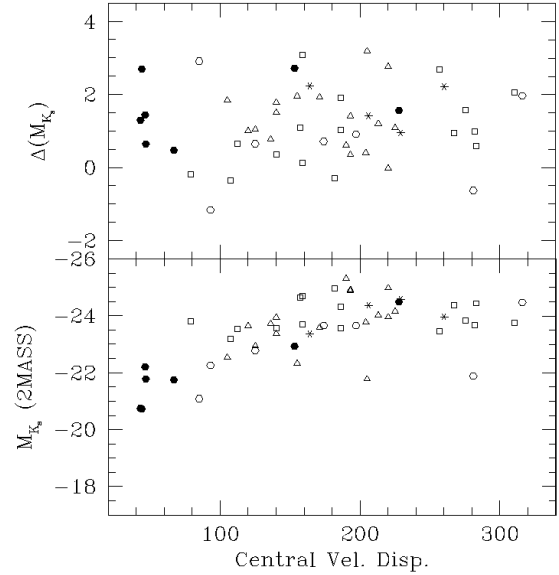}
\end{center}
\caption{$M_{\rm K_s}$ and $\Delta M_{\rm K_s}$ versus central velocity
  dispersion.}\label{fig:v_sig} 
\end{figure}

The lack of correlation with bulge photometric properties make it
intriguing that the absolute magnitude does correlate with the central
velocity dispersion, which is primarily a property of the galaxy's
bulge (see Figure~\ref{fig:v_sig}).  Presumably, this correlation is a
manifestation of the well-known conspiracy by which rotation curves
remain fairly flat across regions dominated by bulge, disk or halo:
this conspiracy means that the bulge dispersion will be tightly
related to the rotation speed in the disk, and, as we have seen above,
disk properties do correlate with total luminosity.

In our attempt to understand the origins of the offset from the spiral
galaxy TFR, it is the correlation of parameters with this offset,
$\Delta M_{K_s}$, that is of more immediate interest.  For most plots,
however, there are no significant correlations to be seen; where there is some
hint of a correlation, it goes in the opposite sense to that seen with
$M_{K_s}$, suggesting that it may well be the kind of artifact
discussed at the beginning of this section.  

\begin{figure}
\begin{center}
\includegraphics[scale=0.4]{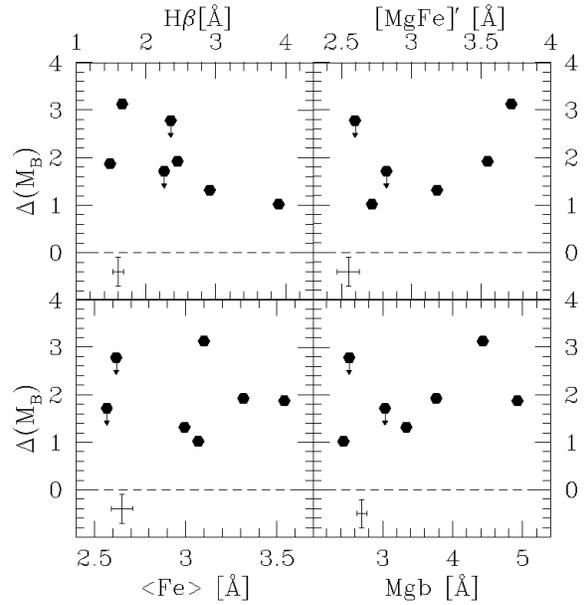}
\end{center}
\caption{$\Delta M_{\rm B}$ versus $H\beta$, $\langle Fe\rangle$, $[MgFe]'$
  and $Mgb$ central line indices for our Fornax data. The error bar in the
  lower left corner of each panel shows the median error. Dots with arrows
  indicate upper limits in $\Delta M_{\rm B}$.}\label{fig:v_index}
\end{figure}

\subsubsection{Spectral parameters}
A further hint as to the origins of the scatter in the TFR can be
found by considering line index measurements and the infered
luminosity-weighted ages from stellar populations synthesis models. For the
reasons outlined in Section~\ref{sec:data}, from this point we focus our
analysis on our Fornax sample only.  

 Giving the homogeneity of the Fornax dataset, we feel confident to use the
 offset from the $B$-band TFR, $\Delta M_{\rm B}$, for the rest of the
 analysis. Despite the fact that uncertainties in RC3 $B$-band photometry are
 larger than in the 2MASS $K_s$-band counterpart, the total errors in $\Delta
 M_{\rm B}$ are dominated by the uncertainties of the TFR from spirals. Also,
 the $B$-band data is more sensitive to recent changes in the star formation
 history than the infrared band, being more suitable for this kind of
 study. In fact, similar results are found if the analysis is performed using
 $K_s$-band data.

Figure~\ref{fig:v_index} shows $\Delta
M_{\rm B}$ plotted against different central indices ($H\beta$, $Mgb$) and the
combined indices $\langle Fe \rangle$ and $[MgFe]'$. The only
indication of a trend with the offset from the TFR comes from the $H\beta$
index, where $\Delta M_{\rm B}$ seems to get weaker as $H\beta$ increases, and stronger as $H\beta$ decreases.  This correlation is
interesting since it is in the sense expected if the degree of fading of S0s
is driven by the timescale since the last significant star formation.

\begin{figure*}
\begin{center}
\includegraphics[scale=0.9]{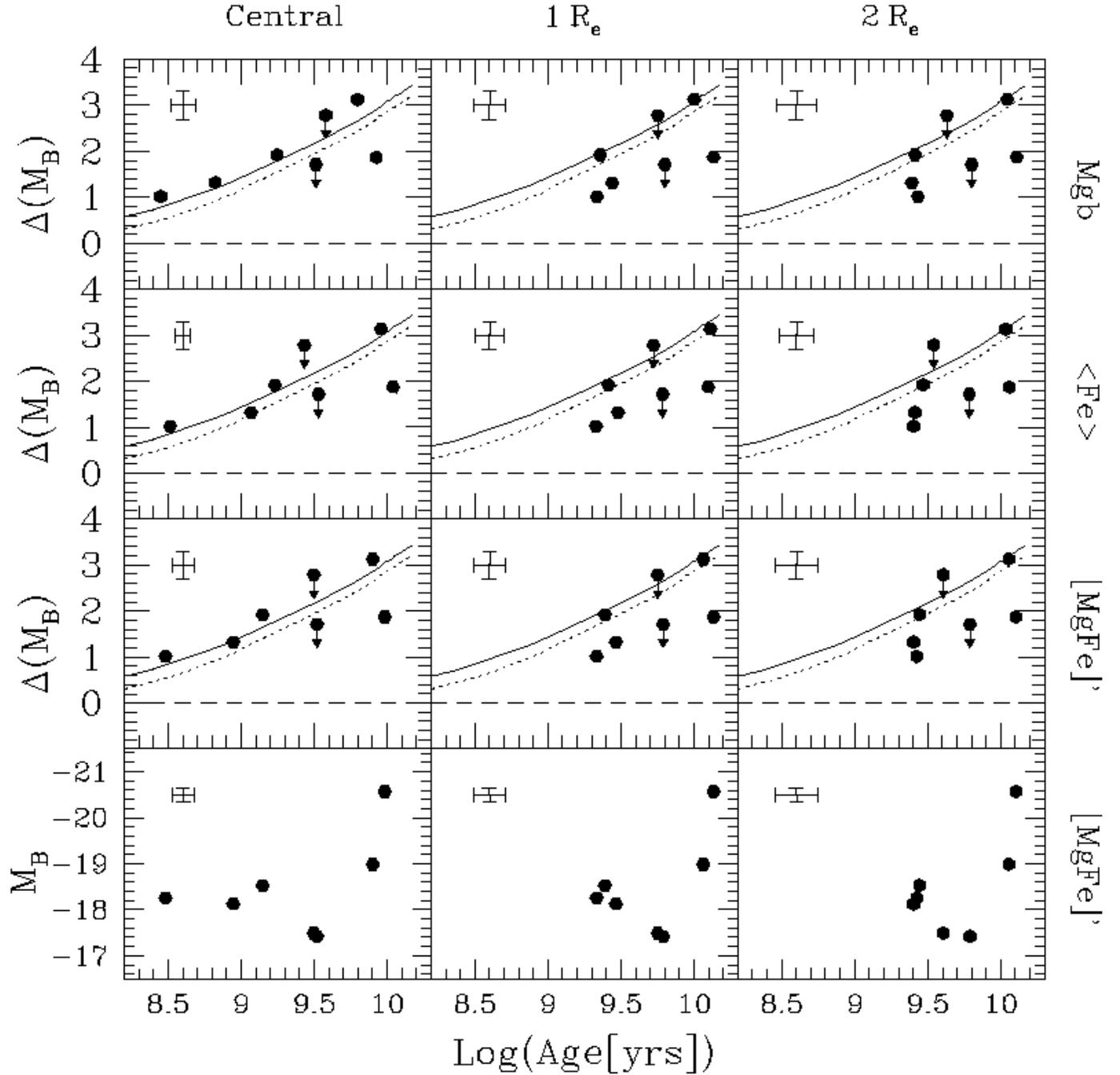}
\end{center}
\caption{\small $\Delta M_{\rm B}$ and $M_{\rm B}$ versus ages in the center,
  at $1\,R_e$ and at $2\,R_e$ of the bulge, using different metallicity
  indicators ($Mgb$, $\langle Fe \rangle$ and $[MgFe]'$). BC03 predictions of
  fading vs. age are plotted for the ``truncation'' (solid line) and
  ``starburst'' (dotted line) models described in section 3.1.2. Dashed
  horizontal lines represent the $B$-band TFR of spirals from Tully \& Pierce
  (2000). The error bar in the upper left corner of each panel shows the
  median error. Dots with arrows indicate upper limits in $\Delta M_{\rm
  B}$. For ESO\,358$-$G059 (dot with arrow of smaller shift) we re-plot the
  results at $1\,R_e$ in the column corresponding to $2\,R_e$.}
  \label{fig:Mb_age}
\end{figure*}

This possibility can be examined more quantitatively if we use the age we
derived from these spectral indices
in Section~\ref{sec:data}.  To ascertain whether the resulting
ages have being significantly influenced by enhancement of
$\alpha$-elements via the well-known age-metallicity degeneracy, we consider
the estimates derived using $Mgb$, $\langle Fe \rangle$ and $[MgFe]'$
parameters as the metallicity-sensitive index; the results are shown in
Figure~\ref{fig:Mb_age}.  The first column of this figure confirms the trend
indicated by the central line indices alone: the fading of the Fornax Cluster
S0s seems to be accompanied by an ageing of their stellar population.
Suggestively, the size of the effect is also of the amplitude expected
for a passively fading stellar population: as we saw in
Section~\ref{sec:shiftint}, over $\sim 1.6\,{\rm Gyr}$, we would expect
a stellar population to fade by $\sim 1.7$ magnitudes in the
$B$-band, just as seen with these data. The predictions from these models
are overplotted in all the pertinent panels of Figure~\ref{fig:Mb_age}. The
second and third columns of this figure present similar data at $1\,R_e$ and
$2\,R_e$. The correlations are stronger for central values than at larger
galactocentric distances: a Spearman ranking correlation coefficient test
shows that the correlation between central age and $\Delta M_{\rm B}$ is
significant at between the 90\% to 97.5\% level, while at larger radii the
confidence level decreases to a range between 85\% to 90\% level only. Apart
from the obvious increment in age uncertainties with S/N ratio (so with
radius), there is another plausible explanation for the apparent radial
dependence of the relation. If high redshift spirals simply swith-off their
star formation in order to become local S0s, the age versus $\Delta M_{\rm B}$
relation would be sensitive to the relative differences in luminosity-weighted
age at the moment the passive fading starts. The importance of these
differences will depend, among other factors, on the relative SFRs of the
progenitor spirals and their radial SFR profile before truncation. This would
introduce additional scatter to the discussed result. However, there is a
possible mechanism which would diminish this effect in the inner regions: a
central starburst. If a central starburst precedes the truncation of the star
formation, the central ages would be ``set to zero'' just before passive
fading starts. In consequence, relative differences in central age are
minimised and the age versus $\Delta M_{\rm B}$ relation becomes tighter. A
central starburst has been proposed in the past as a possible (even as a
necessary) step in the morphological transformation of a spiral into a S0
galaxy (Shioya et al. 2004, Christlein \& Zabludoff 2004) and there is
observational evidence in favour of this scenario (Poggianti et al. 2001;
Mehlert et al. 2000, 2003; Moss \& Whittle 2000). The youngest galaxies of our
Fornax sample seem to have clear positive gradients in age, in agreement with
this hypothesis. However, it will be in Paper III where we will be able to
address this issue in detail, when we perform a detailed study of the stellar
populations of these systems. Also worthy of note in Figure~\ref{fig:Mb_age}
is the absence of a similar correlation between age and $M_{B}$ at all radii.
First, this lack of correlation rules out the possibility that the correlation
with $\Delta M_{B}$ could be spuriously driven by a correlation with
$M_{B}$, as discussed above.  Second and more interestingly, it
implies that there is no compelling evidence for the ``downsizing'' scenario
(Gavazzi 1993, Boselli et al. 2001, Scodeggio et al. 2002) in S0s, which would
predict that the faintest galaxies should have the youngest stellar
populations.  
\\
Although suggestive, these results are not yet definitive. The significance
level of the age versus $\Delta M_{B}$ relation is not strong enough to
totally discard the null hypothesis, and in case of a positive correlation,
many questions remain open: does this relation depends on local environment;
do all S0s follow this trend or other properties, like luminosity, must be
taken into account?. Clearly, what is needed is a larger set of high-quality
data across a wide range of luminosities and environments to finally
disentangle the life histories of these surprisingly complex objects. 

\section{Conclusions}\label{sec:conc}
This paper presents the largest compilation of data to-date in a study
of the Tully--Fisher relation of nearby S0 galaxies in both $B$ and
$K_s$-bands, using new and archival results for a sample of 60
galaxies in different environments.  The principal results of this
study are:

\begin{itemize}
  
\item The local TFR of S0 galaxies presents a shift with respect to
  the local TFR of spirals. Interpreted as a shift in luminosity, it
  amounts to between $-1.3 \pm 0.1$ and $-1.7 \pm 0.4$ in the $B$-band
  and between $-0.8 \pm 0.4$ and $-1.2 \pm 0.4 $ in $K_s$-band, with
  the value depending on the calibration adopted for the spiral galaxy
  TFR.
  
\item In a scenario where S0 galaxies are the descendants of spirals
  at higher redshifts which have simply passively faded, these offsets
  are consistent with a population that has been passively fading for
  somewhere between 1 and 6 Gyr (again depending on the adopted
  calibration for the spiral galaxy TFR).  If we include a starburst
  before the truncation of the star formation, the upper end of the
  possible timescale since transformation increases somewhat.
  However, the large scatter in the TFR means that such a simple model
  with a single epoch of formation cannot provide the whole story.
  
\item The total scatter in the TFR of S0s is found to be $0.88 \pm
  0.06$ mag in the $B$-band and $0.98 \pm 0.06$ in the $K_s$-band.
  Our fitting indicates that only $\sim 10\%$ of this scatter can be
  attributed to the observational errors, with $\sim 90\%$ arising
  from the intrinsic astrophysical spread in the TFR.  Such a scatter
  could have been ``imprinted'' by the larger scatter in the spiral
  galaxy TFR at high redshift, or it could have arisen through the
  subsequent somewhat stochastic evolution of these systems.

\item To try to distinguish between these possibilities, we have
  looked for correlations between the offset from the TFR and other
  properties of these systems.  Morphologically, a number of S0
  properties correlate with their absolute magnitudes, as seen in other
  galaxies, but none seem to correlate strongly with the offset of this
  magnitude from the spiral galaxy TFR.
  
\item By investigating line indices, we have been able to establish a
  reasonably strong correlation between the central ages of the Fornax Cluster
  galaxies' stellar populations and their offsets from the TFR.  This
  offset is in the sense and of the magnitude expected for a
  population of galaxies that transformed from spirals at various
  times in the past and has been passively fading ever since. A central
  starburst before the truncation of star formation would explain why the
  correlation is stronger with central age than at larger galactocentric
  distances.

\end{itemize}

The new data from the Fornax Cluster show that we are finally in a
position to use observations of nearby S0 galaxies to obtain important
archaeological clues as to the mechanisms by which these systems form.
However, the relatively small number of S0s in a single cluster will
always limit the significance of the resulting measurements.  Clearly,
we need a still-larger homogeneous dataset of the quality now
attainable with telescopes like the VLT of S0 galaxies to finish off the work
begun here.

\section*{Acknowledgments}

We would like to thank Dr. O. Nakamura, Dr. S. Bamford,
Dr. M. Mouhcine, Dr. N. Cardiel, Dr. Jes\'us Falc\'on-Barroso and
Dr. Ignacio Trujillo for their help, suggestions and interesting
discussion.

This work was based on observations made with ESO telescopes at Paranal
Observatory under programme ID 070.A-0332. This publication makes use of data
products from the Two Micron All Sky Survey, which is a joint project of the
University of Massachusetts and the Infrared Processing and Analysis
Center/California Institute of Technology, funded by the National
Aeronautics and Space Administration and the National Science
Foundation.

\appendix

\section{Tables}

In this appendix we include tables with different parameters for each galaxy
of our combined sample.\\

Table~A1 includes a compilation of absolute magnitudes (total $K_s$-band from
2MASS) and shifts in magnitude from the $K_s$-band TFR of spiral galaxies,
log($V_{max}$) and central velocity dispersions. Also, in the same table we
present different structural parameters estimated using $K_s$-band photometry
and GIM2D software (Simard et. al. 2002).\\

In Table~A2 we present absolute magnitudes (total $B$-band from RC3) and
shifts in magnitude from the $B$-band TFR of spiral galaxies, central line
indices (Lick/IDS system, measured at 3$\AA$ resolution within $R_e/8$) and
luminosity-weighted ages and metallicities for our sample of 7 S0 galaxies in
Fornax cluster. Ages and metallicities were estimated using Bruzual \& Charlot
(2003) simple stellar population models. Similar data at $1$ and $2\,R_e$ of
the bulge are presented in Table~A3.

\begin{landscape}
\begin{table}
\begin{center}
\caption{Tully-Fisher and Structural Parameters of S0 galaxies.}
\begin{tabular}{@{}clccccccccccc@{}}
\hline
\hline
 $\#$ & Name & $K_{s,tot}$ & $M_{\rm K_s}$ & $\Delta (M_{\rm K_s})$ & $Log(V_{max})$ & $\sigma_0$ & $B/T$ & $R_{\rm e}$ & $R_{\rm d}$ & $i$ & S\'ersic & $R_{\rm half}$ \\
     &   & [mag] & [mag] & [mag] & [$\rm km\,s^{-1}$] & [$\rm km\,s^{-1}$] &  & [Kpc] &  [Kpc]  & [${}^o$] &   &  [Kpc]  \\
    (1)   &      (2) &     (3)  &   (4)  &    (5)    &     (6)    &   (7)   & (8) & (9) & (10) & (11) & (12) & (13)\\
\hline
\hline
FIELD & (N99)$^{1}$ \\
\hline
 1  & NGC\,0584 & 7.30 (0.02) & -24.16 (0.10) &    1.09 (0.26)&  2.40 (0.09)& 225  (--)$^{a}$ & $0.75_{\phantom{1}-0.02}^{\phantom{1}+0.06}$ & $1.50_{\phantom{1}-0.04}^{\phantom{1}+0.22}$ & $3.25_{\phantom{1}-0.34}^{\phantom{1}+0.34}$ & $57.5_{\phantom{1}-3.7}^{\phantom{1}+4.4}$ & $3.83_{\phantom{1}-0.08}^{\phantom{1}+0.18}$ &  2.27 \\\\ 
 2  & NGC\,0936 & 6.91 (0.02) & -24.92 (0.15) &    1.42 (0.28)&  2.52 (0.13)& 193  (--)$^{a}$ & $0.38_{\phantom{1}-0.00}^{\phantom{1}+0.00}$ & $1.36_{\phantom{1}-0.02}^{\phantom{1}+0.02}$ & $4.36_{\phantom{1}-0.02}^{\phantom{1}+0.01}$ & $33.9_{\phantom{1}-0.2}^{\phantom{1}+0.4}$ & $2.92_{\phantom{1}-0.02}^{\phantom{1}+0.03}$ &  4.62 \\\\ 
 3  & NGC\,1023 & 6.24 (0.02) & -24.02 (0.16) &    1.21 (0.29)&  2.40 (0.03)& 213  (--)$^{a}$ & $0.48_{\phantom{1}-0.00}^{\phantom{1}+0.00}$ & $0.86_{\phantom{1}-0.00}^{\phantom{1}+0.01}$ & $3.12_{\phantom{1}-0.02}^{\phantom{1}+0.01}$ & $74.3_{\phantom{1}-0.1}^{\phantom{1}+0.1}$ & $3.70_{\phantom{1}-0.02}^{\phantom{1}+0.04}$ &  2.69 \\\\ 
 4  & NGC\,1052 & 7.45 (0.01) & -23.78 (0.22) &    0.41 (0.33)&  2.28 (0.09)& 204  (--)$^{a}$ & $0.73_{\phantom{1}-0.02}^{\phantom{1}+0.01}$ & $1.39_{\phantom{1}-0.05}^{\phantom{1}+0.04}$ & $1.53_{\phantom{1}-0.05}^{\phantom{1}+0.04}$ & $48.2_{\phantom{1}-2.7}^{\phantom{1}+1.5}$ & $4.29_{\phantom{1}-0.05}^{\phantom{1}+0.05}$ &  1.77 \\\\ 
 5  & NGC\,2549 & 8.05 (0.01) & -22.33 (0.23) &    1.96 (0.34)&  2.29 (0.07)& 155  (--)$^{a}$ & $0.58_{\phantom{1}-0.01}^{\phantom{1}+0.01}$ & $0.48_{\phantom{1}-0.01}^{\phantom{1}+0.01}$ & $1.46_{\phantom{1}-0.01}^{\phantom{1}+0.02}$ & $75.7_{\phantom{1}-0.3}^{\phantom{1}+0.2}$ & $2.85_{\phantom{1}-0.03}^{\phantom{1}+0.02}$ &  1.05 \\\\ 
 6  & NGC\,2768 & 7.00 (0.03) & -24.88 (0.18) &    0.35 (0.30)&  2.40 (0.07)& 193  (--)$^{a}$ & $0.52_{\phantom{1}-0.01}^{\phantom{1}+0.00}$ & $3.12_{\phantom{1}-0.05}^{\phantom{1}+0.02}$ & $5.23_{\phantom{1}-0.04}^{\phantom{1}+0.04}$ & $70.6_{\phantom{1}-0.3}^{\phantom{1}+0.3}$ & $3.98_{\phantom{1}-0.03}^{\phantom{1}+0.05}$ &  3.52 \\\\ 
 7  & NGC\,2787 & 7.26 (0.01) & -21.78 (0.34) &    3.19 (0.41)&  2.37 (0.06)& 205  (--)$^{a}$ & $0.54_{\phantom{1}-0.08}^{\phantom{1}+0.07}$ & $0.34_{\phantom{1}-0.06}^{\phantom{1}+0.05}$ & $0.89_{\phantom{1}-0.07}^{\phantom{1}+0.07}$ & $54.1_{\phantom{1}-2.3}^{\phantom{1}+2.4}$ & $2.43_{\phantom{1}-0.31}^{\phantom{1}+0.23}$ &  0.72 \\\\ 
 8  & NGC\,3115 & 5.88 (0.02) & -23.95 (0.11) &    2.76 (0.26)&  2.57 (0.07)& 220  (--)$^{a}$ & $0.59_{\phantom{1}-0.01}^{\phantom{1}+0.00}$ & $0.84_{\phantom{1}-0.02}^{\phantom{1}+0.01}$ & $2.22_{\phantom{1}-0.00}^{\phantom{1}+0.00}$ & $68.2_{\phantom{1}-0.1}^{\phantom{1}+0.1}$ & $2.98_{\phantom{1}-0.06}^{\phantom{1}+0.01}$ &  1.72 \\\\ 
 9  & NGC\,3384 & 6.75 (0.02) & -23.37 (0.15) &    1.79 (0.28)&  2.39 (0.06)& 140  (--)$^{a}$ & $0.51_{\phantom{1}-0.05}^{\phantom{1}+0.04}$ & $0.42_{\phantom{1}-0.07}^{\phantom{1}+0.06}$ & $2.45_{\phantom{1}-0.20}^{\phantom{1}+0.13}$ & $63.7_{\phantom{1}-2.1}^{\phantom{1}+1.7}$ & $3.16_{\phantom{1}-0.52}^{\phantom{1}+0.48}$ &  1.53 \\\\ 
 10 & NGC\,3412 & 7.67 (0.01) & -22.54 (0.12) &    1.84 (0.27)&  2.30 (0.08)& 105  (--)$^{a}$ & $0.43_{\phantom{1}-0.02}^{\phantom{1}+0.01}$ & $0.39_{\phantom{1}-0.02}^{\phantom{1}+0.01}$ & $1.72_{\phantom{1}-0.07}^{\phantom{1}+0.04}$ & $62.4_{\phantom{1}-0.7}^{\phantom{1}+0.8}$ & $3.60_{\phantom{1}-0.16}^{\phantom{1}+0.14}$ &  1.58 \\\\ 
 11 & NGC\,3489 & 7.37 (0.01) & -22.94 (0.15) &    1.04 (0.28)&  2.26 (0.12)& 125  (--)$^{a}$ & $0.66_{\phantom{1}-0.02}^{\phantom{1}+0.01}$ & $0.55_{\phantom{1}-0.02}^{\phantom{1}+0.02}$ & $1.44_{\phantom{1}-0.04}^{\phantom{1}+0.03}$ & $65.8_{\phantom{1}-0.8}^{\phantom{1}+0.6}$ & $3.64_{\phantom{1}-0.10}^{\phantom{1}+0.08}$ &  1.01 \\\\ 
 12 & NGC\,4251 & 7.73 (0.01) & -23.65 (0.18) &    1.01 (0.30)&  2.33 (0.08)& 120  (--)$^{a}$ & $0.77_{\phantom{1}-0.01}^{\phantom{1}+0.01}$ & $1.34_{\phantom{1}-0.03}^{\phantom{1}+0.02}$ & $1.58_{\phantom{1}-0.04}^{\phantom{1}+0.06}$ & $78.1_{\phantom{1}-0.4}^{\phantom{1}+0.5}$ & $3.64_{\phantom{1}-0.11}^{\phantom{1}+0.06}$ &  1.67 \\\\ 
 13 & NGC\,4382 & 6.14 (0.02) & -25.32 (0.14) &    0.61 (0.28)&  2.48 (0.06)& 190  (--)$^{a}$ & $0.37_{\phantom{1}-0.00}^{\phantom{1}+0.00}$ & $2.10_{\phantom{1}-0.01}^{\phantom{1}+0.03}$ & $4.60_{\phantom{1}-0.01}^{\phantom{1}+0.01}$ & $46.5_{\phantom{1}-0.1}^{\phantom{1}+0.1}$ & $4.20_{\phantom{1}-0.02}^{\phantom{1}+0.01}$ &  5.65 \\\\ 
%14 &NGC\,4649  & 5.74 (0.02) & -25.24 (0.14) &  1.57 (0.28)&  2.58 (0.12)&      (  ) & ---- & ---- & ---- & ---- & ---- & ---- \\\\ 
 14 &NGC\,4753  & 6.72 (0.02) & -24.98 (0.15) &   -0.02 (0.28)&  2.37 (0.04)& 220  (--)$^{a}$ & $0.51_{\phantom{1}-0.00}^{\phantom{1}+0.01}$ & $1.90_{\phantom{1}-0.01}^{\phantom{1}+0.03}$ & $3.69_{\phantom{1}-0.03}^{\phantom{1}+0.03}$ & $59.6_{\phantom{1}-0.2}^{\phantom{1}+0.3}$ & $4.18_{\phantom{1}-0.02}^{\phantom{1}+0.04}$ &  4.00 \\\\ 
 15 &NGC\,4754  & 7.41 (0.03) & -23.60 (0.14) &  1.93 (0.28)&  2.43 (0.07)& 171  (--)$^{a}$ & --- & --- & --- & --- & --- & --- \\\\ 
 16 &NGC\,5866  & 6.87 (0.02) & -23.95 (0.10) &    1.50 (0.26)&  2.42 (0.04)& 140  (--)$^{a}$ & $0.88_{\phantom{1}-0.05}^{\phantom{1}+0.01}$ & $2.20_{\phantom{1}-0.28}^{\phantom{1}+0.04}$ & $1.66_{\phantom{1}-0.09}^{\phantom{1}+0.22}$ & $84.6_{\phantom{1}-1.3}^{\phantom{1}+0.8}$ & $2.75_{\phantom{1}-0.05}^{\phantom{1}+0.02}$ &  2.30 \\\\ 
 17 &NGC\,7332  & 8.01 (0.01) & -23.72 (0.17) &    0.77 (0.30)&  2.31 (0.03)& 136  (--)$^{a}$ & $0.39_{\phantom{1}-0.01}^{\phantom{1}+0.01}$ & $0.46_{\phantom{1}-0.02}^{\phantom{1}+0.02}$ & $2.16_{\phantom{1}-0.03}^{\phantom{1}+0.04}$ & $77.7_{\phantom{1}-0.2}^{\phantom{1}+0.3}$ & $3.23_{\phantom{1}-0.04}^{\phantom{1}+0.07}$ &  2.05 \\ 
\hline
\end{tabular}\\
\end{center}
\footnotesize{Notes: For all pertinent calculations,
  $H_0=70\,$km$\,$s$^{-1}\,$Mpc$^{-1}\,$. All structural parameters derived
  from $K_s$-band 2MASS photometry using GIM2D software (Simard et al. 2002). From (3) to (7), $1\,\sigma$ rms errors
  between ``()''; from (8) to (12), $99\%$ confidence intervals are presented; Col (1), number of each galaxy in our combined sample; Col
  (2), name; Col (3), total apparent magnitude in $K_s$-band using 2MASS
  photometry; Col (4), absolute magnitude in $K_s$-band, using redshifts/distance modulus described in main text; Col (5), shift in
  $K_s$-band magnitude from Tully-Fisher relation of spiral galaxies from
  Tully \& Pierce (2000); Col (6), logarithm (base 10) of the Maximum
  Rotational Velocity as published by the different authors of each subsample; Col (7), central
  velocity dispersion; Col (8), bulge to total fraction; Col (9), effective
  radius of the bulge component; Col (10), exponential disk scale length; Col (11), inclination angle;
  Col (12), S\'ersic index; Col (13), half-light radius (computed by numerical integration of the best structural parameters).\\ ($^{1}$) Individual
  distance modules from Tonry et al. (2001) were used.\\ References: ($^{a}$)
  Neistein et al. (1999); ($^{b}$) Hinz et al. (2001); ($^{c}$) Bernardi et
  al. (2002); ($^{d}$) Our own data; ($^{e}$) Prugniel \& Simien (1995),
  unpublished measurements from OHP; ($^{f}$) di Nella et al. (1995).}
\end{table}
\end{landscape}

\begin{landscape}
\begin{table}
\begin{center}
\begin{tabular}{@{}clccccccccccc@{}}
\hline
\hline
 $\#$ & Name & $K_{s,tot}$ & $M_{\rm K_s}$ & $\Delta (M_{\rm K_s})$ & $Log(V_{max})$ & $\sigma_0$ & $B/T$ & $R_{\rm e}$ & $R_{\rm d}$ & $i$ & S\'ersic & $R_{\rm half}$ \\
     &   & [mag] & [mag] & [mag] & [$\rm km\,s^{-1}$] & [$\rm km\,s^{-1}$] &  & [Kpc] &  [Kpc]  & [${}^o$] &   &  [Kpc]  \\
    (1)   &      (2) &     (3)  &   (4)  &    (5)    &     (6)    &   (7)   & (8) & (9) & (10) & (11) & (12) & (13)\\
\hline
\hline
COMA$^{2}$ \\
\hline
  18 &  DOI\,215       & 11.12 (0.04) & -23.83 (0.05) &  1.58 (0.24)& 2.42 (0.05)& 275.4 ( -- )$^{b}$ & --- & --- & --- & --- & --- & --- \\\\ 
  19 &  IC\,3943       & 11.27 (0.04) & -23.68 (0.05) &  1.00 (0.24)& 2.33 (0.03)& 281.8 ( -- )$^{b}$ & --- & --- & --- & --- & --- & --- \\\\ 
  20 &  IC\,3990       & 10.50 (0.03) & -24.44 (0.04) &  0.58 (0.24)& 2.37 (0.04)& 282.9 ( -- )$^{b}$ & --- & --- & --- & --- & --- & --- \\\\ 
  21 &  NGC\,4787      & 11.40 (0.06) & -23.55 (0.06) &  0.66 (0.25)& 2.28 (0.03)& 112.2 ( -- )$^{b}$ & $0.91_{\phantom{1}-0.07}^{\phantom{1}+0.05}$ & $6.53_{\phantom{1}-1.06}^{\phantom{1}+2.13}$ & $1.26_{\phantom{1}-0.40}^{\phantom{1}+0.65}$ & $64.7_{\phantom{1}-5.1}^{\phantom{1}+5.1}$ & $3.62_{\phantom{1}-0.16}^{\phantom{1}+0.24}$ & 5.60 \\\\  
  22 &  NGC\,4892      & 10.63 (0.03) & -24.32 (0.04) &  1.02 (0.24)& 2.41 (0.01)& 186.1 ( -- )$^{b}$ & $0.14_{\phantom{1}-0.03}^{\phantom{1}+0.05}$ & $1.33_{\phantom{1}-0.37}^{\phantom{1}+0.65}$ & $4.56_{\phantom{1}-0.22}^{\phantom{1}+0.19}$ & $77.4_{\phantom{1}-0.8}^{\phantom{1}+0.7}$ & $3.21_{\phantom{1}-0.45}^{\phantom{1}+0.09}$ & 6.71 \\\\ 
  23 &  NGC\,4931      & 10.31 (0.02) & -24.64 (0.04) &  1.09 (0.24)& 2.46 (0.02)& 157.3 ( -- )$^{b}$ & $0.78_{\phantom{1}-0.16}^{\phantom{1}+0.04}$ & $3.41_{\phantom{1}-1.31}^{\phantom{1}+0.29}$ & $9.08_{\phantom{1}-2.79}^{\phantom{1}+1.66}$ & $47.0_{\phantom{1}-8.6}^{\phantom{1}+13.5}$ & $3.96_{\phantom{1}-0.60}^{\phantom{1}+0.24}$ & 5.12 \\\\
  24 &  NGC\,4934      & 11.15 (0.04) & -23.80 (0.05) & -0.19 (0.24)& 2.21 (0.03)&  79.2 ( -- )$^{b}$ & $0.35_{\phantom{1}-0.06}^{\phantom{1}+0.05}$ & $1.88_{\phantom{1}-0.40}^{\phantom{1}+0.34}$ & $2.86_{\phantom{1}-0.14}^{\phantom{1}+0.13}$ & $81.7_{\phantom{1}-0.8}^{\phantom{1}+0.8}$ & $3.67_{\phantom{1}-0.32}^{\phantom{1}+0.30}$ & 3.84 \\\\ 
  25 &  NGC\,4944      & 10.00 (0.03) & -24.95 (0.04) & -0.30 (0.24)& 2.33 (0.02)& 182.1 ( -- )$^{b}$ & $0.57_{\phantom{1}-0.07}^{\phantom{1}+0.08}$ & $2.95_{\phantom{1}-0.40}^{\phantom{1}+0.49}$ & $6.17_{\phantom{1}-0.49}^{\phantom{1}+0.38}$ & $75.9_{\phantom{1}-1.8}^{\phantom{1}+1.5}$ & $3.00_{\phantom{1}-0.06}^{\phantom{1}+0.05}$ & 5.59 \\\\ 
  26 &  NGC\,4966      & 10.25 (0.03) & -24.70 (0.04) &  0.12 (0.24)& 2.35 (0.04)& 159.0 ( -- )$^{b}$ & $0.39_{\phantom{1}-0.04}^{\phantom{1}+0.02}$ & $1.25_{\phantom{1}-0.23}^{\phantom{1}+0.17}$ & $4.20_{\phantom{1}-0.25}^{\phantom{1}+0.15}$ & $58.4_{\phantom{1}-1.9}^{\phantom{1}+1.4}$ & $3.66_{\phantom{1}-0.18}^{\phantom{1}+0.10}$ & 4.47 \\\\ 
  27 &  ZW\,160$-$034  & 11.76 (0.05) & -23.18 (0.06) & -0.36 (0.25)& 2.12 (0.04)& 107.2 ( -- )$^{b}$ & --- & --- & --- & --- & --- & --- \\\\ 
  28 &  ZW\,160$-$083  & 11.40 (0.05) & -23.55 (0.06) &  0.35 (0.25)& 2.25 (0.06)& 140.2 ( -- )$^{b}$ & --- & --- & --- & --- & --- & --- \\\\ 
  29 &  ZW\,160$-$101  & 11.20 (0.04) & -23.75 (0.05) &  2.06 (0.24)& 2.46 (0.02)& 310.6 ( -- )$^{b}$ & $0.68_{\phantom{1}-0.04}^{\phantom{1}+0.05}$ & $1.29_{\phantom{1}-0.09}^{\phantom{1}+0.15}$ & $3.46_{\phantom{1}-0.52}^{\phantom{1}+0.43}$ & $80.2_{\phantom{1}-2.2}^{\phantom{1}+1.8}$ & $2.41_{\phantom{1}-0.21}^{\phantom{1}+0.26}$ & 2.11 \\\\ 
  30 &  ZW\,160$-$107  & 10.59 (0.03) & -24.36 (0.04) &  0.94 (0.24)& 2.40 (0.01)& 267.2 ( -- )$^{b}$ & $0.41_{\phantom{1}-0.02}^{\phantom{1}+0.02}$ & $1.33_{\phantom{1}-0.13}^{\phantom{1}+0.12}$ & $2.82_{\phantom{1}-0.10}^{\phantom{1}+0.09}$ & $81.1_{\phantom{1}-0.6}^{\phantom{1}+0.4}$ & $3.35_{\phantom{1}-0.32}^{\phantom{1}+0.19}$ & 3.26 \\\\ 
  31 &  ZW\,160$-$109  & 11.13 (0.05) & -23.82 (0.06) &  1.50 (0.25)& 2.41 (0.09)&  --   ( -- ) & $0.75_{\phantom{1}-0.20}^{\phantom{1}+0.10}$ & $2.26_{\phantom{1}-0.80}^{\phantom{1}+0.38}$ & $2.67_{\phantom{1}-0.46}^{\phantom{1}+0.35}$ & $47.9_{\phantom{1}-10.1}^{\phantom{1}+7.9}$ & $3.28_{\phantom{1}-0.19}^{\phantom{1}+0.38}$ & 2.84 \\\\
  32 &  ZW\,160$-$119  & 11.38 (0.04) & -23.57 (0.05) &  0.96 (0.24)& 2.32 (0.04)&  --   ( -- ) & --- & --- & --- & --- & --- & --- \\\\ 
  33 &  ZW\,160$-$214  & 11.04 (0.04) & -23.91 (0.05) &  1.07 (0.24)& 2.37 (0.06)&  --   ( -- ) & $0.71_{\phantom{1}-0.08}^{\phantom{1}+0.07}$ & $1.49_{\phantom{1}-0.21}^{\phantom{1}+0.19}$ & $4.07_{\phantom{1}-0.72}^{\phantom{1}+0.34}$ & $72.4_{\phantom{1}-3.9}^{\phantom{1}+3.2}$ & $2.38_{\phantom{1}-0.23}^{\phantom{1}+0.28}$ & 2.34 \\\\ 
  34 &  IC\,3946       & 11.12 (0.04) & -23.83 (0.05) &  1.66 (0.24)& 2.43 (0.03)&  --   ( -- ) & $0.79_{\phantom{1}-0.09}^{\phantom{1}+0.08}$ & $1.59_{\phantom{1}-0.25}^{\phantom{1}+0.36}$ & $3.78_{\phantom{1}-0.72}^{\phantom{1}+0.30}$ & $76.9_{\phantom{1}-5.1}^{\phantom{1}+3.8}$ & $3.11_{\phantom{1}-0.24}^{\phantom{1}+0.31}$ & 2.21 \\\\ 
  35 &  IC\,3955       & 11.38 (0.05) & -23.57 (0.06) &  1.91 (0.25)& 2.43 (0.07)& 186.0 ( 9.0)$^{c}$ & $0.46_{\phantom{1}-0.13}^{\phantom{1}+0.10}$ & $1.40_{\phantom{1}-0.62}^{\phantom{1}+0.48}$ & $2.64_{\phantom{1}-0.22}^{\phantom{1}+0.32}$ & $61.4_{\phantom{1}-5.2}^{\phantom{1}+4.4}$ & $4.41_{\phantom{1}-0.61}^{\phantom{1}+1.04}$ & 3.07 \\\\ 
 36  &  IC\,3976       & 11.50 (0.04) & -23.45 (0.05) &  2.69 (0.24)& 2.50 (0.04)& 257.0 ( 6.0)$^{c}$ & --- & --- & --- & --- & --- & --- \\\\ 
 37  &  IC\,4111       & 12.13 (0.09) & -22.82 (0.10) & -0.84 (0.26)& 2.03 (0.04)&  --   ( -- ) & $0.57_{\phantom{1}-0.13}^{\phantom{1}+0.10}$ & $3.26_{\phantom{1}-0.60}^{\phantom{1}+0.53}$ & $2.75_{\phantom{1}-0.40}^{\phantom{1}+0.50}$ & $76.2_{\phantom{1}-4.3}^{\phantom{1}+4.4}$ & $3.27_{\phantom{1}-0.63}^{\phantom{1}+0.39}$ & 3.95 \\\\ 
 38  &  NGC\,4873      & 11.25 (0.04) & -23.70 (0.06) &  3.08 (0.25)& 2.57 (0.06)& 159.0 (11.0)$^{c}$ & --- & --- & --- & --- & --- & ---  \\\\ 
 39  &  UGC\,8122      & 11.45 (0.06) & -23.50 (0.07) &  0.42 (0.25)& 2.25 (0.15)&  --   ( -- ) & $0.34_{\phantom{1}-0.11}^{\phantom{1}+0.11}$ & $2.55_{\phantom{1}-0.56}^{\phantom{1}+0.34}$ & $3.32_{\phantom{1}-0.41}^{\phantom{1}+0.41}$ & $51.5_{\phantom{1}-7.9}^{\phantom{1}+4.6}$ & $3.20_{\phantom{1}-0.47}^{\phantom{1}+0.72}$ & 4.63 \\ 
\hline		  
\end{tabular}\\
\end{center}
\footnotesize{($^{2}$) Redshift of 0.0227 for Coma cluster assumed (Smith et
  al. 2004)}
\end{table}
\end{landscape}

\begin{landscape}
\begin{table}
\begin{center}
\begin{tabular}{@{}clccccccccccc@{}}
\hline
\hline
 $\#$ & Name & $K_{s,tot}$ & $M_{\rm K_s}$ & $\Delta (M_{\rm K_s})$ & $Log(V_{max})$ & $\sigma_0$ & $B/T$ & $R_{\rm e}$ & $R_{\rm d}$ & $i$ & S\'ersic & $R_{\rm half}$ \\
     &   & [mag] & [mag] & [mag] & [$\rm km\,s^{-1}$] & [$\rm km\,s^{-1}$] &  & [Kpc] &  [Kpc]  & [${}^o$] &   &  [Kpc]  \\
    (1)   &      (2) &     (3)  &   (4)  &    (5)    &     (6)    &   (7)   & (8) & (9) & (10) & (11) & (12) & (13)\\
\hline
\hline
 FORNAX$^{3}$ \\
\hline
  40  &  NGC\,1380       &  6.87 (0.02) & -24.49 (0.07) & 1.56 (0.25)& 2.49 (0.04)& 227.8 ( 1.8)$^{d}$ & $0.58_{\phantom{1}-0.01}^{\phantom{1}+0.01}$ & $1.59_{\phantom{1}-0.03}^{\phantom{1}+0.03}$ & $3.28_{\phantom{1}-0.04}^{\phantom{1}+0.09}$ & $66.9_{\phantom{1}-0.4}^{\phantom{1}+0.6}$ & $3.34_{\phantom{1}-0.03}^{\phantom{1}+0.03}$ &  3.04 \\\\
  41  &  NGC\,1381  	 &  8.42 (0.02) & -22.93 (0.07) & 2.72 (0.25)& 2.44 (0.05)& 153.1 ( 1.8)$^{d}$ & $0.57_{\phantom{1}-0.01}^{\phantom{1}+0.01}$ & $0.67_{\phantom{1}-0.01}^{\phantom{1}+0.02}$ & $1.88_{\phantom{1}-0.05}^{\phantom{1}+0.07}$ & $82.5_{\phantom{1}-0.3}^{\phantom{1}+0.4}$ & $3.07_{\phantom{1}-0.10}^{\phantom{1}+0.14}$ &  1.48 \\\\
  42  &  NGC\,1380A 	 &  9.57 (0.04) & -21.78 (0.08) & 0.64 (0.25)& 2.08 (0.01)&  46.8 ( 0.9)$^{d}$ & $0.18_{\phantom{1}-0.03}^{\phantom{1}+0.06}$ & $0.98_{\phantom{1}-0.25}^{\phantom{1}+0.32}$ & $1.80_{\phantom{1}-0.06}^{\phantom{1}+0.04}$ & $78.1_{\phantom{1}-0.8}^{\phantom{1}+0.6}$ & $3.71_{\phantom{1}-0.10}^{\phantom{1}+0.07}$ &  2.69 \\\\
  43  &  NGC\,1375       &  9.61 (0.03) & -21.75 (0.08) & 0.47 (0.25)& 2.06 (0.08)&  66.8 ( 1.3)$^{d}$ & $0.19_{\phantom{1}-0.02}^{\phantom{1}+0.01}$ & $0.29_{\phantom{1}-0.03}^{\phantom{1}+0.02}$ & $1.33_{\phantom{1}-0.07}^{\phantom{1}+0.04}$ & $67.5_{\phantom{1}-1.2}^{\phantom{1}+0.8}$ & $2.22_{\phantom{1}-0.08}^{\phantom{1}+0.07}$ &  1.79 \\\\
  44  &  IC\,1963  	 &  9.15 (0.02) & -22.20 (0.07) & 1.44 (0.25)& 2.22 (0.04)&  46.3 ( 1.4)$^{d}$ & $0.39_{\phantom{1}-0.03}^{\phantom{1}+0.03}$ & $2.04_{\phantom{1}-0.12}^{\phantom{1}+0.21}$ & $1.42_{\phantom{1}-0.02}^{\phantom{1}+0.03}$ & $84.6_{\phantom{1}-0.2}^{\phantom{1}+0.2}$ & $3.61_{\phantom{1}-0.09}^{\phantom{1}+0.10}$ &  2.29 \\\\
  45  &  ESO\,358$-$G006 & 10.62 (0.05) & -20.73 (0.09) & 2.70 (0.26)& 2.19 (0.07)&  43.9 ( 0.9)$^{d}$ & --- & --- & --- & --- & --- & --- \\\\           		          		           		        	              											             
  46  &  ESO\,358$-$G059 & 10.60 (0.04) & -20.75 (0.08) & 1.30 (0.25)& 2.04 (0.22)&  42.9 ( 1.0)$^{d}$ & $0.72_{\phantom{1}-0.07}^{\phantom{1}+0.04}$ & $1.46_{\phantom{1}-0.14}^{\phantom{1}+0.20}$ & $0.21_{\phantom{1}-0.02}^{\phantom{1}+0.01}$ & $61.9_{\phantom{1}-2.0}^{\phantom{1}+2.0}$ & $2.06_{\phantom{1}-0.40}^{\phantom{1}+0.17}$ & 0.91 \\
\hline
 VIRGO$^{4}$ \\
\hline
 47 & NGC\,4352 & 9.87 (0.04) & -21.08 (0.04) &  2.91 (0.24)& 2.26 (0.03)&  85.0 ( 8.0)$^{c}$ & $0.76_{\phantom{1}-0.09}^{\phantom{1}+0.07}$ & $1.08_{\phantom{1}-0.17}^{\phantom{1}+0.13}$ & $2.02_{\phantom{1}-0.30}^{\phantom{1}+0.23}$ & $80.1_{\phantom{1}-5.4}^{\phantom{1}+3.4}$ & $3.40_{\phantom{1}-0.20}^{\phantom{1}+0.24}$ & 1.52\\\\ 
 48 & NGC\,4417 & 8.17 (0.03) & -22.78 (0.03) &  0.65 (0.24)& 2.19 (0.05)& 125.0 ( 4.0)$^{c}$ & $0.56_{\phantom{1}-0.02}^{\phantom{1}+0.02}$ & $0.62_{\phantom{1}-0.02}^{\phantom{1}+0.03}$ & $1.82_{\phantom{1}-0.05}^{\phantom{1}+0.04}$ & $78.2_{\phantom{1}-0.4}^{\phantom{1}+0.4}$ & $2.79_{\phantom{1}-0.06}^{\phantom{1}+0.07}$ & 1.37\\\\
 49 & NGC\,4435 & 7.30 (0.02) & -23.66 (0.01) &  0.72 (0.24)& 2.30 (0.03)& 174.0 (16.0)$^{c}$ & $0.53_{\phantom{1}-0.04}^{\phantom{1}+0.10}$ & $1.32_{\phantom{1}-0.15}^{\phantom{1}+0.19}$ & $4.20_{\phantom{1}-0.13}^{\phantom{1}+0.00}$ & $15.9_{\phantom{1}-11.1}^{\phantom{1}+33.3}$ & $4.18_{\phantom{1}-0.44}^{\phantom{1}+0.17}$ & 3.53\\\\
 50 & NGC\,4442 & 7.29 (0.02) & -23.66 (0.02) &  0.91 (0.24)& 2.32 (0.02)& 197.0 (15.0)$^{c}$ & $0.67_{\phantom{1}-0.01}^{\phantom{1}+0.02}$ & $0.92_{\phantom{1}-0.02}^{\phantom{1}+0.04}$ & $2.62_{\phantom{1}-0.05}^{\phantom{1}+0.06}$ & $73.5_{\phantom{1}-0.5}^{\phantom{1}+0.8}$ & $3.03_{\phantom{1}-0.03}^{\phantom{1}+0.06}$ & 1.63\\\\
 51 & NGC\,4452 & 9.07 (0.03) & -21.88 (0.03) & -0.63 (0.24)& 1.94 (0.02)& 281.0 (15.0)$^{e}$ & $0.21_{\phantom{1}-0.07}^{\phantom{1}+0.04}$ & $1.37_{\phantom{1}-0.40}^{\phantom{1}+0.24}$ & $1.27_{\phantom{1}-0.03}^{\phantom{1}+0.02}$ & $85.2_{\phantom{1}-0.3}^{\phantom{1}+0.2}$ & $2.99_{\phantom{1}-0.10}^{\phantom{1}+0.11}$ & 2.00\\\\
 52 & NGC\,4474 & 8.70 (0.02) & -22.26 (0.02) & -1.16 (0.24)& 1.93 (0.04)&  93.0 ( 7.0)$^{c}$ & $0.79_{\phantom{1}-0.01}^{\phantom{1}+0.02}$ & $1.21_{\phantom{1}-0.07}^{\phantom{1}+0.07}$ & $1.34_{\phantom{1}-0.05}^{\phantom{1}+0.09}$ & $81.2_{\phantom{1}-0.6}^{\phantom{1}+1.0}$ & $4.04_{\phantom{1}-0.41}^{\phantom{1}+0.26}$ & 1.47\\\\
 53 & NGC\,4526 & 6.47 (0.02) & -24.48 (0.02) &  1.96 (0.24)& 2.54 (0.01)& 316.0 ( 7.0)$^{c}$ & $0.61_{\phantom{1}-0.01}^{\phantom{1}+0.01}$ & $1.37_{\phantom{1}-0.02}^{\phantom{1}+0.02}$ & $4.25_{\phantom{1}-0.02}^{\phantom{1}+0.04}$ & $75.9_{\phantom{1}-0.3}^{\phantom{1}+0.3}$ & $2.96_{\phantom{1}-0.04}^{\phantom{1}+0.05}$ & 2.80\\\\
 54 & NGC\,4638 & 8.21 (0.02) & -22.74 (0.02) &  1.85 (0.24)& 2.32 (0.02)&  --   ( -- ) & $0.74_{\phantom{1}-0.03}^{\phantom{1}+0.01}$ & $1.29_{\phantom{1}-0.22}^{\phantom{1}+0.08}$ & $0.50_{\phantom{1}-0.01}^{\phantom{1}+0.04}$ & $79.1_{\phantom{1}-0.5}^{\phantom{1}+0.6}$ & $3.74_{\phantom{1}-0.10}^{\phantom{1}+0.15}$ & 1.09\\
\hline
FIELD & (M02)$^{5}$ \\
\hline
 55 &   NGC\,1184 & 8.12 (0.02) & -24.59 (0.02) & 0.95 (0.24)& 2.43 (0.03)& 229.0 (14.0)$^{f}$ & $0.31_{\phantom{1}-0.01}^{\phantom{1}+0.01}$ & $1.07_{\phantom{1}-2.43}^{\phantom{1}+0.07}$ & $3.80_{\phantom{1}-0.05}^{\phantom{1}+0.06}$ & $80.8_{\phantom{1}-0.2}^{\phantom{1}+0.2}$ & $3.58_{\phantom{1}-0.05}^{\phantom{1}+0.07}$ & 4.53 \\\\ 
 56 &   NGC\,1611 & 9.41 (0.01) & -24.53 (0.01) & 1.01 (0.24)& 2.43 (0.02)&  --   ( -- ) & --- & --- & --- & --- & --- & --- \\\\  
 57 &   NGC\,2612 & 8.78 (0.02) & -23.06 (0.02) & 2.20 (0.24)& 2.40 (0.02)&  --   ( -- ) & $0.44_{\phantom{1}-0.01}^{\phantom{1}+0.01}$ & $0.71_{\phantom{1}-2.40}^{\phantom{1}+0.02}$ & $1.83_{\phantom{1}-0.06}^{\phantom{1}+0.07}$ & $75.0_{\phantom{1}-0.6}^{\phantom{1}+0.8}$ & $3.13_{\phantom{1}-0.06}^{\phantom{1}+0.04}$ & 1.88 \\\\ 
 58 &   NGC\,3986 & 8.98 (0.02) & -24.37 (0.02) & 1.41 (0.24)& 2.46 (0.01)& 206.0 (22.0)$^{f}$ & $0.53_{\phantom{1}-0.01}^{\phantom{1}+0.02}$ & $1.70_{\phantom{1}-2.46}^{\phantom{1}+0.09}$ & $5.67_{\phantom{1}-0.11}^{\phantom{1}+0.15}$ & $84.0_{\phantom{1}-0.2}^{\phantom{1}+0.3}$ & $2.48_{\phantom{1}-0.06}^{\phantom{1}+0.09}$ & 4.13 \\\\ 
 59 &   NGC\,4179 & 7.92 (0.02) & -23.36 (0.04) & 2.24 (0.24)& 2.44 (0.01)& 164.0 (15.0)$^{c}$ & $0.59_{\phantom{1}-0.01}^{\phantom{1}+0.01}$ & $0.83_{\phantom{1}-2.44}^{\phantom{1}+0.02}$ & $2.20_{\phantom{1}-0.04}^{\phantom{1}+0.05}$ & $80.5_{\phantom{1}-0.2}^{\phantom{1}+0.3}$ & $3.41_{\phantom{1}-0.07}^{\phantom{1}+0.06}$ & 1.73 \\\\ 
 60 &   NGC\,5308 & 8.36 (0.03) & -23.97 (0.05) & 2.22 (0.24)& 2.51 (0.02)& 260.0 (12.0)$^{f}$ & $0.39_{\phantom{1}-0.01}^{\phantom{1}+0.04}$ & $0.62_{\phantom{1}-2.51}^{\phantom{1}+0.11}$ & $2.68_{\phantom{1}-0.05}^{\phantom{1}+0.12}$ & $81.6_{\phantom{1}-0.2}^{\phantom{1}+0.5}$ & $2.99_{\phantom{1}-0.06}^{\phantom{1}+0.04}$ & 2.59 \\ 
\hline
%\begin{tabular}{@{}lccccccccccc@{}}
\end{tabular}\\
\end{center}
\footnotesize{($^{3}$) Distance modulus of 31.35 mag for Fornax cluster
  assumed (Madore et al. 1999). ($^{4}$) Individual redshifts were used
  from RC3. ($^{5}$) Individual redshidts were used from NASA/IPAC
  Extragalactic Database (NED); for NGC\,1611 and NGC\,2612, redshifts were
  estimated from spectra by M02.}
\end{table}
\end{landscape}

\begin{landscape}
\begin{table}
\begin{center}
\caption{Central Line Indices, Ages, Metallicities and $\Delta (M_{\rm B})$ of S0 galaxies in Fornax.}
\begin{tabular}{@{}lcccccccccccc@{}}
\hline
\hline
 Name & $M_{\rm B}$ & $\Delta (M_{\rm B})$ & $H\beta$ & $Mgb$ & $\langle Fe
\rangle$ & $[MgFe]'$ & $Age_{Mgb}$ & $[Fe/H]_{Mgb}$ & $Age_{\langle Fe \rangle}$ & $[Fe/H]_{\langle Fe \rangle}$ & $Age_{[MgFe]'}$ & $[Fe/H]_{[MgFe]'}$ \\
     & [mag] & [mag] &[\AA] & [\AA] & [\AA] & [\AA] & [Gyrs] &  [dex]  & [Gyrs] &
  [dex] & [Gyrs] & [dex]\\

    (1)   &      (2) &     (3)  &   (4)  &    (5)    &     (6)    &   (7)   &
          (8) & (9) & (10) & (11) & (12) & (13)  \\

\hline
\hline
CENTRAL \\
\hline
      NGC\,1380& -20.57 (0.12)& 1.87 (0.27) & 1.48 (0.04) & 4.92 (0.04)& 3.54 (0.03)& 4.15 (0.05)&$ 8.4_{\phantom{1}-0.9}^{\phantom{1}+1.0}$ &$0.70_{\phantom{1}-0.06}^{\phantom{1}+0.07}$&$10.9_{\phantom{1}-1.3}^{\phantom{1}+1.4}$& $   0.37_{\phantom{1}-0.04}^{\phantom{1}+0.04}$& $9.6_{\phantom{1}-1.4}^{\phantom{1}+1.5}$ & $0.49_{\phantom{1}-0.06}^{\phantom{1}+0.06}$ \\\\
      NGC\,1381& -18.98 (0.12)& 3.13 (0.27) & 1.66 (0.04) & 4.43 (0.04)& 3.10 (0.03)& 3.72 (0.05)&$ 6.2_{\phantom{1}-1.0}^{\phantom{1}+1.1}$ &$0.56_{\phantom{1}-0.08}^{\phantom{1}+0.08}$&$ 9.0_{\phantom{1}-0.9}^{\phantom{1}+1.1}$& $   0.16_{\phantom{1}-0.04}^{\phantom{1}+0.04}$& $7.9_{\phantom{1}-0.8}^{\phantom{1}+1.0}$ & $0.32_{\phantom{1}-0.06}^{\phantom{1}+0.06}$ \\\\
     NGC\,1380A& -18.12 (0.15)& 1.32 (0.28) & 2.91 (0.08) & 3.34 (0.08)& 2.99 (0.07)& 3.18 (0.09)&$ 0.7_{\phantom{1}-0.1}^{\phantom{1}+0.2}$ &$1.35_{\phantom{1}-0.19}^{\phantom{1}+0.20}$&$ 1.2_{\phantom{1}-0.1}^{\phantom{1}+0.1}$& $   0.65_{\phantom{1}-0.06}^{\phantom{1}+0.06}$& $0.9_{\phantom{1}-0.2}^{\phantom{1}+0.2}$ & $0.96_{\phantom{1}-0.13}^{\phantom{1}+0.13}$ \\\\
      NGC\,1375& -18.25 (0.15)& 1.02 (0.28) & 3.90 (0.09) & 2.44 (0.08)& 3.07 (0.06)& 2.72 (0.09)&$ 0.3_{\phantom{1}-0.1}^{\phantom{1}+0.1}$ &$1.46_{\phantom{1}-0.20}^{\phantom{1}+0.21}$&$ 0.3_{\phantom{1}-0.05}^{\phantom{1}+0.05}$& $ 1.00_{\phantom{1}-0.05}^{\phantom{1}+0.05}$& $0.3_{\phantom{1}-0.06}^{\phantom{1}+0.1}$& $1.24_{\phantom{1}-0.13}^{\phantom{1}+0.14}$ \\\\
       IC\,1963& -18.53 (0.12)& 1.92 (0.27) & 2.45 (0.08) & 3.76 (0.07)& 3.32 (0.06)& 3.55 (0.08)&$ 1.8_{\phantom{1}-0.04}^{\phantom{1}+0.1}$ &$1.09_{\phantom{1}-0.14}^{\phantom{1}+0.14}$&$ 1.7_{\phantom{1}-0.2}^{\phantom{1}+0.2}$& $   0.69_{\phantom{1}-0.05}^{\phantom{1}+0.05}$& $1.4_{\phantom{1}-0.2}^{\phantom{1}+0.2}$ & $0.92_{\phantom{1}-0.11}^{\phantom{1}+0.11}$ \\\\
\scriptsize{ESO\,358$-$G006}& -17.49 (0.16)& 2.78 (0.29) & 2.35 (0.15) & 2.52 (0.14)& 2.62 (0.12)& 2.60 (0.16)&$ 3.8_{\phantom{1}-1.6}^{\phantom{1}+2.1}$ &$-0.25_{\phantom{1}-0.16}^{\phantom{1}+0.18}$&$2.7_{\phantom{1}-0.5}^{\phantom{1}+1.0}$& $   0.10_{\phantom{1}-0.11}^{\phantom{1}+0.12}$& $3.1_{\phantom{1}-0.8}^{\phantom{1}+1.8}$ & $-0.07_{\phantom{1}-0.17}^{\phantom{1}+0.18}$ \\\\
\scriptsize{ESO\,358$-$G059}& -17.41 (0.15)& 1.71 (0.28) & 2.26 (0.07) & 3.03 (0.06)& 2.57 (0.05)& 2.82 (0.07)&$ 3.2_{\phantom{1}-0.5}^{\phantom{1}+0.8}$ &$0.07_{\phantom{1}-0.09}^{\phantom{1}+0.10}$&$ 3.4_{\phantom{1}-0.5}^{\phantom{1}+0.6}$& $   0.02_{\phantom{1}-0.05}^{\phantom{1}+0.05}$& $3.3_{\phantom{1}-0.5}^{\phantom{1}+0.7}$ & $0.05_{\phantom{1}-0.08}^{\phantom{1}+0.08}$ \\
\hline
\end{tabular}\\
\end{center}
\footnotesize{Notes: From (2) to (7), $1\,\sigma$ rms errors
  between ``()''; from (8) to (13), $99\%$ confidence intervals are
  presented. Col (1), name; Col (2), absolute total $B$-band magnitude from
  RC3 assuming distance modulus of 31.35 mag (Madore et al. 1999) for Fornax cluster; Col (3), shift in
  $B$-band magnitude from Tully-Fisher relation of spiral galaxies from
  Tully \& Pierce (2000); Col (4), $H\beta$ index; Col (5), $Mgb$ index; Col (6),
  $\langle Fe \rangle$ combined index (Gorgas, Efstathiou \&
  Arag\'on-Salamanca 1990); Col (7), $[MgFe]'$ combined index (Gonz\'alez
  1993; Thomas, Maraston \& Bender 2003); Col (8) (9), age and metallicity, estimated using Bruzual \& Charlot
  (2003) simple stellar population models and $Mgb$ as metallicity indicator;
  Col (10) (11), age and metallicity, estimated using $\langle Fe \rangle$ as
  metallicity indicator; Col (12) (13), age and metallicity, estimated using $[MgFe]'$ as metallicity indicator.
  }
\end{table}
\end{landscape}

\begin{landscape}
\begin{table}
\begin{center}
\caption{Line Indices, Ages and Metallicities at $1$ and $2\,R_e$ of S0 galaxies in Fornax.}
\begin{tabular}{@{}lcccccccccc@{}}
\hline
\hline
 Name & $H\beta$ & $Mgb$ & $\langle Fe \rangle$ & $[MgFe]'$ & $Age_{Mgb}$ & $[Fe/H]_{Mgb}$ & $Age_{\langle Fe \rangle}$ & $[Fe/H]_{\langle Fe \rangle}$ & $Age_{[MgFe]'}$ & $[Fe/H]_{[MgFe]'}$ \\
     &[\AA] & [\AA] & [\AA] & [\AA] & [Gyrs] &  [dex]  & [Gyrs] &  [dex] &
 [Gyrs] & [dex]\\

    (1)   &      (2) &     (3)  &   (4)  &    (5)    &     (6)    &   (7)   &
          (8) & (9) & (10) & (11)  \\

\hline
\hline
At $1\,R_e$ \\
\hline 
      NGC\,1380&  1.61 (0.08) & 3.62 (0.07)& 2.86 (0.06)& 3.21 (0.08)&$13.6_{\phantom{1}-3.0}^{\phantom{1}+4.7}$ &$-0.13_{\phantom{1}-0.08}^{\phantom{1}+0.09}$&$12.4_{\phantom{1}-2.4}^{\phantom{1}+3.3}$&$-0.06_{\phantom{1}-0.07}^{\phantom{1}+0.07}$&$13.5_{\phantom{1}-2.9}^{\phantom{1}+4.6}$ & $-0.14_{\phantom{1}-0.09}^{\phantom{1}+0.09}$ \\\\
      NGC\,1381&  1.61 (0.08) & 3.98 (0.07)& 2.82 (0.06)& 3.36 (0.09)&$10.0_{\phantom{1}-2.1}^{\phantom{1}+3.0}$ &$ 0.15_{\phantom{1}-0.10}^{\phantom{1}+0.11}$&$12.9_{\phantom{1}-2.7}^{\phantom{1}+4.4}$&$-0.11_{\phantom{1}-0.08}^{\phantom{1}+0.07}$&$11.5_{\phantom{1}-2.5}^{\phantom{1}+3.6}$ & $0.00_{\phantom{1}-0.10}^{\phantom{1}+0.10}$ \\\\
     NGC\,1380A&  2.26 (0.11) & 3.24 (0.10)& 2.72 (0.08)& 2.98 (0.12)&$ 2.7_{\phantom{1}-0.5}^{\phantom{1}+1.0}$ &$ 0.26_{\phantom{1}-0.16}^{\phantom{1}+0.17}$&$3.0_{\phantom{1}-0.5}^{\phantom{1}+0.9}$& $0.14_{\phantom{1}-0.08}^{\phantom{1}+0.08}$& $2.9_{\phantom{1}-0.6}^{\phantom{1}+1.0}$ & $0.19_{\phantom{1}-0.13}^{\phantom{1}+0.14}$ \\\\
      NGC\,1375&  2.50 (0.15) & 2.94 (0.14)& 2.75 (0.12)& 2.86 (0.16)&$ 2.2_{\phantom{1}-0.2}^{\phantom{1}+0.6}$ &$ 0.26_{\phantom{1}-0.20}^{\phantom{1}+0.22}$&$2.1_{\phantom{1}-0.1}^{\phantom{1}+0.4}$& $0.28_{\phantom{1}-0.12}^{\phantom{1}+0.13}$& $2.1_{\phantom{1}-0.1}^{\phantom{1}+0.5}$ & $0.27_{\phantom{1}-0.19}^{\phantom{1}+0.20}$ \\\\
       IC\,1963&  2.18 (0.08) & 3.74 (0.08)& 3.19 (0.06)& 3.45 (0.08)&$ 2.3_{\phantom{1}-0.3}^{\phantom{1}+0.5}$ &$ 0.66_{\phantom{1}-0.14}^{\phantom{1}+0.15}$&$2.6_{\phantom{1}-0.3}^{\phantom{1}+0.4}$& $0.47_{\phantom{1}-0.06}^{\phantom{1}+0.06}$& $2.4_{\phantom{1}-0.3}^{\phantom{1}+0.5}$ & $0.55_{\phantom{1}-0.11}^{\phantom{1}+0.11}$ \\\\
ESO\,358$-$G006&  2.15 (0.15) & 2.77 (0.14)& 2.40 (0.12)& 2.57 (0.16)&$ 5.6_{\phantom{1}-1.5}^{\phantom{1}+2.3}$ &$-0.24_{\phantom{1}-0.15}^{\phantom{1}+0.16}$&$5.3_{\phantom{1}-1.3}^{\phantom{1}+1.9}$& $-0.15_{\phantom{1}-0.12}^{\phantom{1}+0.12}$&$5.6_{\phantom{1}-1.4}^{\phantom{1}+2.2}$ & $-0.24_{\phantom{1}-0.16}^{\phantom{1}+0.17}$ \\\\
ESO\,358$-$G059&  2.08 (0.16) & 2.82 (0.14)& 2.38 (0.12)& 2.64 (0.17)&$ 6.3_{\phantom{1}-1.8}^{\phantom{1}+2.9}$ &$-0.26_{\phantom{1}-0.16}^{\phantom{1}+0.17}$&$6.1_{\phantom{1}-1.6}^{\phantom{1}+2.5}$& $-0.21_{\phantom{1}-0.13}^{\phantom{1}+0.13}$&$6.1_{\phantom{1}-1.7}^{\phantom{1}+2.7}$ & $-0.22_{\phantom{1}-0.16}^{\phantom{1}+0.18}$ \\
\hline
At $2\,R_e$ \\
\hline
      NGC\,1380&   1.65 (0.10) & 3.49 (0.10)& 2.83 (0.08)& 3.12 (0.11)&$ 12.8_{\phantom{1}-3.5}^{\phantom{1}+6.6}$&$-0.18_{\phantom{1}-0.11}^{\phantom{1}+0.12}$&$11.4_{\phantom{1}-2.5}^{\phantom{1}+3.0}$&$-0.07_{\phantom{1}-0.09}^{\phantom{1}+0.09}$& $12.6_{\phantom{1}-3.4}^{\phantom{1}+6.4}$ &$-0.17_{\phantom{1}-0.12}^{\phantom{1}+0.12}$ \\\\
      NGC\,1381&   1.71 (0.08) & 3.44 (0.08)& 2.70 (0.06)& 3.04 (0.09)&$ 11.0_{\phantom{1}-2.2}^{\phantom{1}+2.5}$&$-0.15_{\phantom{1}-0.09}^{\phantom{1}+0.10}$&$10.8_{\phantom{1}-2.1}^{\phantom{1}+3.4}$&$-0.15_{\phantom{1}-0.08}^{\phantom{1}+0.07}$& $11.3_{\phantom{1}-2.4}^{\phantom{1}+4.0}$ &$-0.19_{\phantom{1}-0.10}^{\phantom{1}+0.10}$ \\\\
     NGC\,1380A&   2.34 (0.09) & 3.28 (0.09)& 2.91 (0.07)& 3.10 (0.10)&$ 2.4_{\phantom{1}-0.4}^{\phantom{1}+0.5}$ &$0.40_{\phantom{1}-0.15}^{\phantom{1}+0.16}$& $2.6_{\phantom{1}-0.3}^{\phantom{1}+0.3}$& $0.35_{\phantom{1}-0.07}^{\phantom{1}+0.07}$&  $2.5_{\phantom{1}-0.4}^{\phantom{1}+0.4}$ & $0.36_{\phantom{1}-0.12}^{\phantom{1}+0.12}$ \\\\
      NGC\,1375&   2.39 (0.16) & 2.82 (0.14)& 2.63 (0.12)& 2.72 (0.16)&$ 2.7_{\phantom{1}-0.6}^{\phantom{1}+1.6}$ &$0.04_{\phantom{1}-0.20}^{\phantom{1}+0.21}$& $2.5_{\phantom{1}-0.4}^{\phantom{1}+0.9}$& $0.13_{\phantom{1}-0.12}^{\phantom{1}+0.12}$&  $2.6_{\phantom{1}-0.5}^{\phantom{1}+1.3}$ & $0.06_{\phantom{1}-0.18}^{\phantom{1}+0.19}$ \\\\
       IC\,1963&   2.14 (0.16) & 3.66 (0.14)& 3.07 (0.12)& 3.36 (0.16)&$ 2.6_{\phantom{1}-0.7}^{\phantom{1}+1.7}$ &$0.53_{\phantom{1}-0.27}^{\phantom{1}+0.28}$& $2.9_{\phantom{1}-0.6}^{\phantom{1}+1.3}$& $0.37_{\phantom{1}-0.11}^{\phantom{1}+0.12}$&  $2.8_{\phantom{1}-0.6}^{\phantom{1}+1.6}$ & $0.44_{\phantom{1}-0.20}^{\phantom{1}+0.22}$ \\\\
ESO\,358$-$G006&   2.30 (0.15) & 2.57 (0.14)& 2.42 (0.12)& 2.49 (0.16)&$ 4.3_{\phantom{1}-1.8}^{\phantom{1}+2.1}$ &$-0.25_{\phantom{1}-0.15}^{\phantom{1}+0.18}$&$3.5_{\phantom{1}-0.9}^{\phantom{1}+1.8}$& $-0.08_{\phantom{1}-0.11}^{\phantom{1}+0.11}$& $4.0_{\phantom{1}-1.6}^{\phantom{1}+2.0}$ & $-0.20_{\phantom{1}-0.16}^{\phantom{1}+0.17}$ \\\\
%ESO\,358$-$G059&   2.08 (0.16) & 2.82 (0.14)& 2.38 (0.12)& 2.64 (0.17)&$ 6.3_{\phantom{1}-1.8}^{\phantom{1}+2.9}$ &$-0.26_{\phantom{1}-0.16}^{\phantom{1}+0.17}$&$6.1_{\phantom{1}-1.6}^{\phantom{1}+2.5}$& $-0.21_{\phantom{1}-0.13}^{\phantom{1}+0.13}$& $6.1_{\phantom{1}-1.7}^{\phantom{1}+2.7}$ & $-0.22_{\phantom{1}-0.16}^{\phantom{1}+0.18}$ \\
ESO\,358$-$G059$^{1}$&   --- (---) & --- (---)& --- (---)& --- (---)& --- & --- & --- & --- & --- & --- \\
\hline
\end{tabular}\\
\end{center}
\footnotesize{Notes: From (2) to (5), $1\,\sigma$ rms errors
  between ``()''; from (6) to (11), $99\%$ confidence intervals are
  presented. Col (1), name; Col (2), $H\beta$ index; Col (3), $Mgb$ index; Col (4),
  $\langle Fe \rangle$ combined index (Gorgas, Efstathiou \&
  Arag\'on-Salamanca 1990); Col (5), $[MgFe]'$ combined index (Gonz\'alez
  1993; Thomas, Maraston \& Bender 2003); Col (6) (7), age and metallicity, estimated using Bruzual \& Charlot
  (2003) simple stellar population models and $Mgb$ as metallicity indicator;
  Col (8) (9), age and metallicity, estimated using $\langle Fe \rangle$ as
  metallicity indicator; Col (10) (11), age and metallicity, estimated using $[MgFe]'$ as metallicity indicator.
  }\\ ($^{1}$) For ESO\,358$-$G059 no data is available at $2\,R_e$ of the
  bulge. In consequence, for this particular galaxy, we re-plot in Figure~\ref{fig:Mb_age} the results at
  $1\,R_e$ in the column corresponding to $2\,R_e$. 
\end{table}
\end{landscape}

\bsp

\label{lastpage}

\end{document}